\documentclass[12pt]{article}
\usepackage{graphicx}
\usepackage{amsbsy}
\input epsf
\begin{document}
\def \beq{\begin{equation}}
\def \eeq{\end{equation}}
\rightline{EFI-00-14} \rightline{astro-ph/0205046}
\bigskip
\centerline{\bf A PROTOTYPE SYSTEM FOR DETECTING THE RADIO-FREQUENCY}
\centerline{\bf PULSE ASSOCIATED WITH COSMIC RAY AIR SHOWERS
\footnote{Submitted to Nucl.~Instr.~Meth.}}
\bigskip
\centerline{Kevin Green,\footnote{Present address: Louis Dreyfus
Corporation, Wilton, CT 06897.} Jonathan
L. Rosner, and Denis A. Suprun} \centerline{\it Enrico Fermi
Institute and Department of Physics} \centerline{\it University
of Chicago, Chicago, IL 60637}
\bigskip
\centerline{J. F. Wilkerson}
\centerline{\it Center for Experimental Nuclear Physics and Astrophysics}
\centerline{\it Department of Physics, University
of Washington, Seattle, WA 98195}
\bigskip

\centerline{\bf ABSTRACT}
\begin{quote}
The development of a system to detect the radio-frequency (RF) pulse associated
with extensive air showers of cosmic rays is described.  This work was
performed at the CASA/MIA array in Utah, with the intention of designing
equipment that can be used in conjunction with the Auger Giant Array.  A small
subset of data (less than 40 out of a total of 600 hours of running time),
taken under low-noise conditions, permitted upper limits to be placed on the
rate and strength of pulses accompanying showers with energies around
$10^{17}$~eV.
\end{quote}

\section{Motivation}

As a result of work in the 1960's and 1970's \cite{Jel,Chac,Allan,Atrash},
some of which continued beyond then (see, e.g., \cite{Agasa,GS,Yak,Gau}),
it appears that air showers of energy 10$^{17}$~eV are accompanied by
radio-frequency (RF) pulses \cite{Ask}, whose properties suggest that they
are due mainly to the separation of positive and negative charges of the shower 
in the Earth's magnetic field \cite{KL,SC}. The most convincing data
were accumulated in the 30--100~MHz frequency range. However,
opinions differ regarding the strength of the pulses, and
atmospheric and ionospheric effects have led to irreproducibility
of results. In particular, there may also be pulses associated
with cosmic-ray-induced atmospheric discharges \cite{RRW,atm}.
There are reports of detection at MHz or sub-MHz frequencies
\cite{Agasa,GS,Yak}, which could be associated with such a
mechanism. Signals above 100~MHz have also been reported \cite{Gau}.

A study was undertaken of the feasibility of equipping the Auger
array \cite{Auger} with the ability to detect such pulses.  The
higher energy of the showers to which the array would be
sensitive could change the parameters of detection.  Before a
design for large-scale RF pulse detection could be produced, it
was necessary to retrace some of the steps of the past 30 years
by searching for the pulses accompanying 10$^{17}$~eV showers,
and by studying some of the factors which led to their irreproducibility in
the past.  RF pulses may be able to provide auxiliary information about 
primary composition and shower height \cite{Allan}.

In this article we describe the prototype activity at the
CASA/MIA site and draw some conclusions from it regarding plans
for the Auger project.  We have not been able to demonstrate
the presence of RF pulses at CASA/MIA, and could only
set upper limits for their intensity.  The upper limits on
the rates of events in which the North-South (or East-West)
projection of the signal pulse was greater than some value 
were established at $R^{up}({\cal E}_{\nu NS}>{\cal E}_\nu^0) 
=0.555/({\cal E}_\nu^0)^2 \ $~h$^{-1}$ and 
$R^{up}({\cal E}_{\nu EW}>{\cal E}_\nu^0)=0.889/
({\cal E}_\nu^0)^2 \ $~h$^{-1}$, respectively, with ${\cal E}_\nu$ being
the electric field strength per unit of frequency, measured in 
$\mu{\rm V}/{\rm m}/{\rm MHz}$.  More concrete plans for RF detection at
Auger must await a prototype at the Auger site which utilizes some of the
lessons learned from the present work.  A preliminary description of this
work was presented in Ref.\ \cite{rh}.

In Sections 2 and 3 we discuss expectations for RF signals and
previous claims of their observation.  Section 4 is devoted to
details of the setup at CASA/MIA, including some of the reasons
for choosing the specific configuration utilized.  Our
preliminary results are given in Section 5, while Section 6 deals
with issues specific to a giant array such as Auger.  We
summarize in Section 7.  Appendix A gives details of the
sensitivity calculation, Appendix B establishes some properties
of simulated pulses, while Appendix C summarizes cost estimates
for an installation at the Auger site.

\section{Expectations}

We briefly summarize some expectations \cite{Allan} for the
characteristic of the RF signal associated with cosmic ray air
showers.

\subsection{Mechanisms of pulse generation}

In the 1950's, R. R. Wilson \cite{RRW} proposed that cosmic rays
could induce the atmosphere to act as a giant spark chamber,
triggering discharges of the ambient field gradient.  This
gradient, normally around 100 V/m, can attain values as high as
10 kV/m during intense thunderstorm activity \cite{Allan}.  Thus,
the mechanism would lead to pulses of greatly varying intensity,
whose correlation with air showers would be difficult to
establish unless field gradients could be monitored over the
whole path of the discharge.

Another mechanism of pulse generation is associated with the
asymmetry in electron and positron yields in showers as a result
of Compton and knock-on processes.  By the end of the shower,
electrons outnumber positrons by about 10--25\%, leading to a
transient of vertically moving negative charge \cite{Ask}.  This is thought
to be the main mechanism for generation of radio-frequency signals from 
showers in solid material such as polar ice \cite{ZHS} or sand \cite{Saltz},
but is probably not the dominant mechanism in the atmosphere.

Still another source of electromagnetic radiation in a cosmic ray
shower involves the separation of positive and negative electric
charges in the Earth's magnetic field.  This is thought to be the
dominant mechanism accounting for atmospheric pulses with frequencies in the
30 -- 100~MHz range, and will be taken as the model for the
signal for which the search was undertaken.

\subsection{Characteristic pulse}

The time profile of a pulse due to charge-separation in the
Earth's magnetic field can be modelled \cite{Allan} by assuming
that the bulk of the shower giving rise to the pulse is
concentrated between an altitude of 10 and 5 km (for a shower of
energy $10^{17}$~eV) and calculating the pulse duration by
comparing the total path lengths between the antenna and the
beginning and the end of the shower.  The rise time of a pulse
from a shower with zero zenith angle observed 200 meters from its
axis is expected to be about 5~ns, followed by a longer decay time
and a still longer recovery time with opposite amplitude (about
100~ns) such that the total DC component is zero.

The radiation from any stage of the shower which is traveling directly
toward the antenna is expected to arrive to the antenna about the same time
as the shower itself. The difference is accounted for by the refraction index
of air. Such an essentially $\delta$-function pulse has the highest-frequency
components in its spectrum.  Showers for which the impact parameters of the
cores are farther from the antenna will have reduced high-frequency components
since the total pulse duration will be longer, approaching several microseconds
for vertically incident showers viewed from the side.

The pulse is expected to grow linearly in amplitude with shower
energy as a result of the increased number of particles emitting
RF energy.  This linear growth assumes coherence of the emitting
particles, which is probably a good assumption for RF wavelengths
of several meters. The greater penetration of the atmosphere by
more energetic showers also leads to an increased RF signal since
the radiating particles are closer to the receiver.  This should
make the pulse amplitude increase more rapidly than linearly with
primary energy. However, this effect is largely offset by the
fact that a greater fraction of such deeply penetrating showers
will have reduced high-frequency components in their pulses, as a
result of the greater apparent time taken by the pulse to build
up to its maximum amplitude at the receiver.  The combination of
the above three effects is expected to lead fortuitously to an
overall linear dependence of pulse amplitude on primary energy
\cite{Allan}.

At extremely high energies, shower particles will even be lost by
collision with the Earth.  This may give rise to a different type
of RF signal but will not be effective in the context of the
charge-separation mechanism considered here. The pulses
associated with charge separation in the Earth's magnetic field
should correspond to radio signals with approximately horizontal
polarization.  (For showers not arriving vertically from directly
overhead there will also be a small vertical polarization component.)

\subsection{RF backgrounds}

Discharges of atmospheric electricity constitute an important
source of background pulses.  These will be detected at random
intervals at a rate which depends strongly on local weather
conditions as well as on ionospheric reflections.  Man-made RF
sources include television and radio stations, police and other
communications services, broad-band sources (such as ignition
noise), and sources within the experiment itself.  (We shall
discuss such sources for the CASA/MIA array presently.)  The
propagation of distant noise sources to the receiver is a strong
function of frequency.  During years of sunspot minima (e.g.,
1995--6), ionospheric propagation on frequencies above 25~MHz is
rare except for ``sporadic-E'' propagation, which can permit
signals to arrive from distances of up to 2000 km via a single
reflection from the ionosphere.  As solar activity increases (e.g.,
subsequently to 1996), consistent daytime propagation over even
greater distances can occur on frequencies up to and beyond 30~MHz.

Galactic noise can be the dominant signal in exceptionally
radio-quiet environments for frequencies in the low VHF (30--100
~MHz) range \cite{Allan}. For higher frequencies in such
environments, thermal receiver noise becomes the dominant
effect.  We shall see that the CASA/MIA site is far from quiet
enough that these effects become limiting.

\section{Some previous observations} \label{s:Prev}

An early proposal involved detection of the ionization produced by air 
showers
via radar \cite{radar,PG}.  The first claim for detection of the
charge-separation mechanism utilized relatively narrow-band
techniques at 44 and 70~MHz \cite{Jel}.  A Soviet group
reported signals at 30~MHz \cite{Sov}, while a University of
Michigan group at the BASJE Cosmic Ray Station on Mt.~Chacaltaya,
Bolivia \cite{Chac} studied pulses in the 40--90~MHz range.  The
collaboration of H. R. Allan at Haverah Park in England
\cite{Allan} studied the dependence of signals on primary energy
$E_p$, perpendicular distance $R$ of closest approach of the
shower core, zenith angle $\theta$, and angle $\alpha$ between
the shower axis and the magnetic field vector. Their results
indicate that the electric field strength per unit of frequency,
${\cal E}_\nu$, could be expressed as \beq \label{eqn:E} {\cal
E}_\nu = 20 \frac{E_p}{10^{17} {\rm~eV}} \sin \alpha \cos \theta
\exp \left( - \frac{R}{R_0(\nu, \theta)} \right)~~~\mu{\rm
V}~{\rm m}^{-1}~ {\rm MHz}^{-1}~~~, \eeq where $R_0$ is an increasing
function of $\theta$, equal (for example) to $(110 \pm 10)$ m for
$\nu = 55$~MHz and $\theta < 35^\circ$.

The Haverah Park observations are consistent with the model
mentioned in Section~2.2 in which the pulse's onset is generated
by the start of the shower at an elevation of about 10 km above
sea level, while its end is associated with the greater total
path length (shower $+$ signal propagation distance) associated
with the shower's absorption about 5 km above sea level. (The
elevation at the CASA/MIA site is about 1450 m above sea level;
the average atmospheric depth is 870 g/cm$^2$ \cite{CASAnim}.)

The Haverah Park observations were subsequently updated to
yield field strengths approximately 12 times weaker than
Eq.~(\ref{eqn:E}) \cite{Atrash}, while observations in the
U.S.S.R. gave field strengths approximately 2.2 times weaker than
(\ref{eqn:E}).  Thus, some question persists about the magnitude
of the effect, serving as an impetus to further measurements if
the RF detection technique is to be employed as part of a new
giant array.

More recent pulse detections include claims for pulses with
components below 500~kHz seen by observers at the AGASA array in
Akeno, Japan \cite{Agasa} and a group working at Yakutsk in
Siberia \cite{Yak}, and claims for pulses at VHF frequencies seen
by groups at the Gran Sasso in Italy \cite{GS} and at Gauhati
University in India \cite{Gau}.  There seems to be no unanimity
regarding the time duration, generation mechanism, or intensity
of these pulses.  Related methods have been used to study
lightning-induced pulses \cite{DSmith}.

\section{CASA/MIA Prototype setup}

\subsection{Description of the CASA/MIA detector}

The Chicago Air Shower Array (CASA) \cite{CASAnim} was originally
constituted as a rectangular grid of $33 \times 33$ stations on
the surface of the desert at Dugway Proving Ground, Dugway,
Utah.  The inter-station spacing is 15~m. A station has four 61~
cm $\times$ 61~cm $\times$ 1.27~cm sheets of plastic scintillator
each viewed by its own photomultiplier tube (PMT).  When a signal
appears on 3 of 4 PMTs in a station, a ``trigger request pulse''
of 5~mA with 5~$\mu$s duration is sent to a central trailer, where
a decision is made on whether to interrogate all stations for a
possible event.  Details of this trigger are described in
Ref.~\cite{CASAnim}.

\begin{figure}
\centerline{\epsfysize = 5 in \epsffile{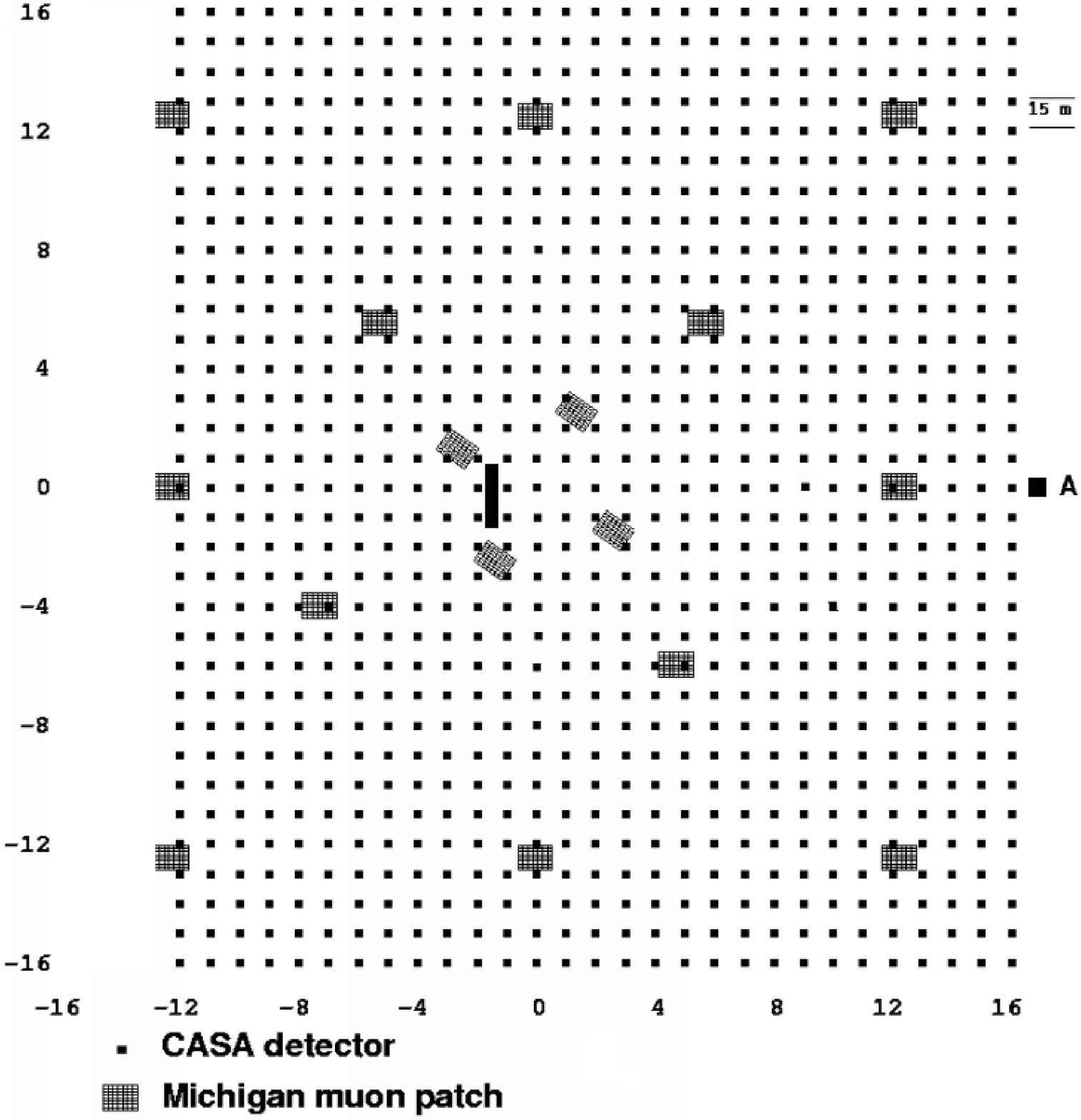}}
\caption{Geometry of the CASA/MIA array.  Small squares denote CASA
stations; cross-hatched rectangles denote muon patches.
Large rectangle near center is the central trailer; rectangle to
right (east) of the array is RF trailer; symbol ``A'' denotes
placement of antenna.}
\end{figure}

For future reference, we shall denote the coordinates of each box
by $(n_x,n_y)$, where $-16 \le (n_x,n_y) \le 16$, $n_x =
(x/15~{\rm m})$, $n_y = (y/15~{\rm m})$, and $(x,y)$ denotes the
position of the center of the box to the (East,~North) of the
center of the array.  When this experiment was begun the CASA
array had already been reconfigured to remove boxes with $-16 \le
n_x \le -13$, i.e., the 4 westernmost ``ribs'' of the array.  For
runs performed in 1998, boxes with $n_x = 16$ had also been
removed from the array.

The University of Michigan collaborators designed and built a muon detection
array (MIA) to operate in conjunction with CASA.  It consists of
sixteen ``patches,'' each having 64 muon counters, buried 3~m
below ground at various locations in the CASA array.  (See
Fig.~1.) Each counter has lateral dimensions 1.9~m $\times$ 1.3~m.
Four of the patches, each about 45~m
from the center of the array, lie on the corners of a skewed
rectangle; four, each about 110~m from the
center of the array, lie on a quadrangle with slightly different
skewed orientation, and eight lie on the
sides and corners of a rectangle with sides $x \simeq \pm 180$~m
and $y \simeq \pm 185$~m.

In April of 1991 the CASA/MIA array was partially disabled by a
lightning strike which hit one of the few trees on the site.  The
array was repaired, and an extensive lightning-protection grid
installed.  The grid consisted of wires strung on poles about 
15~feet above the array, traveling in the $x$, $y$, and $x \pm y$
directions.  This grid turned out to have significant effect on
our choice of parameters for the RF studies.

During the operation of the present experiment, 144 surface
\v{C}erenkov detectors \cite{BLANCA} were distributed throughout
the array.  Other additions to the array, which shall not concern
us, included a stereographic atmospheric \v{C}erenkov
detector (DICE) \cite{DICE} and an optical facility for
communicating with a high-resolution atmospheric fluorescence
detector (HiRes) \cite{HiRes} located on a hilltop several miles away.

\subsection{Initial RF surveys at CASA/MIA site}

In order to determine whether RF pulse detection was feasible at
the CASA site, a spectrum analyzer was used to make a broad
survey of the RF noise at the CASA site in various frequency
ranges and at various locations.  It was determined that in the
central trailer, the broad-band noise associated with various
computers, switching power supplies, and other electronics was so
intense that no RF searches could be undertaken.  The same was
true to a great extent at any position within the perimeter of
the lightning-protection grid.  Moreover, it was deemed unsafe to
erect an antenna above that grid within the perimeter of the
array, since any projecting object would defeat the purpose of
the grid.

Surveys just outside the array indicated a much quieter RF
environment.  An antenna was placed about 24 m east of box
(16,0), corresponding to $x = 263.8$ m, $y=0$ m, and its signal
fed into a trailer located about 10 m closer to the array.  All
further studies were performed using this configuration. (See
Fig.\ 1.)  Nonetheless, there still remained a number of
identifiable noise sources, which we now describe.

\subsubsection{Television and FM broadcast stations} \label{ss:TVFM}

The CASA/MIA site is about 100~km southwest of Salt Lake City, at
first sight affording a reasonably quiet RF environment.
However, many television and FM stations in Salt Lake City
broadcast from a high mountain about 35~km southwest of Salt Lake
City, or 65~km northeast of Dugway.  These are responsible for a
major component of the RF signal in the range which is of
greatest interest to us.  As an example, we summarize the VHF
television stations broadcasting from the above site \cite{TV} in
Table~1.  The video and audio frequencies shown are carrier
frequencies.  Video signals are modulated with vestigial-sideband
modulation, occupying the range from 1.25~MHz below the carrier
frequency to about 3.5~MHz above it.  A color subcarrier lies
3.58~MHz above the video carrier.  Audio signals are frequency
modulated with deviation not exceeding 250~kHz so as to remain
within the total allotted bandwidth of 6~MHz for each channel.
The FM broadcast band, extending from 88 to 108~MHz, is packed
with strong signals, with the strongest typically spaced by the
0.8~MHz interval characteristic of inter-station spacing in a large urban
area.

\begin{table}
\caption{Television stations broadcasting from site 65 km
northeast of Dugway.}
\begin{center}
\begin{tabular}{l c c r r r} \hline
Call sign & Channel &   Band   &  Video &  Audio & Power \\
          &         &   (MHz)  &  (MHz) &  (MHz) &  (kW) \\ \hline
KUTV      & 2       &  54--60  &  55.25 &  59.75 &  45.7 \\
KTVX      & 4       &  66--72  &  67.25 &  71.75 &  32.4 \\
KSL-TV    & 5       &  76--82  &  77.25 &  81.75 &  33.9 \\
KUED      & 7       & 174--180 & 175.25 & 179.75 & 155.0 \\
KULC      & 9       & 186--192 & 187.25 & 191.75 & 166.0 \\
KBYU-TV   & 11      & 198--204 & 199.25 & 203.75 & 162.0 \\
KSTU      & 13      & 210--216 & 211.25 & 215.75 & 112.0 \\ \hline
\end{tabular}
\end{center}
\end{table}

\subsubsection{CASA noise}

The CASA boards contain crystals oscillating at various
frequencies, including 16, 20, and 50~MHz.  The behavior of a
single CASA board was investigated at the University of Chicago.
The various clock signals were detected at short distances ($< 1$
~m) from the board, but a much more intense set of harmonics of 78
~kHz emanated from the switching power supplies.  These harmonics
persisted well above 100~MHz.  At 144--148~MHz (monitored using
an amateur radio transceiver), they overlapped, leading to
intense broad-band noise.

The above signals were considerably less problematic at the RF
trailer. During CASA operation the boards' clock frequencies and
some of their harmonics (including 32, 40, and 48~MHz) were
detectable.  However, noise from the switching power supplies
seemed to be at an acceptably low level.

The CASA boards emit powerful RF pulses when digitizing and
transmitting data. These pulses constituted a major background to
our RF search, and will be discussed in Section 5.  The noise
arrived through the antenna system and not through the trigger
cable or antenna feed cable, as was determined by acquiring data
with a dummy load in place of the antenna.

\subsubsection{Intermittent narrow-band interference}

In addition to persistent RF carriers from TV and FM broadcast
stations, intermittent signals would appear from time to time.
The strongest of these was traced to local narrow-band FM
communications.  This signal was so strong that digital filtering
methods (to be described below) were powerless to eliminate it.
Consequently, any event containing such a signal was discarded
for further analysis.

\subsubsection{Low-frequency interference sources}

Although the majority of survey work dealt with frequencies above
25~MHz, some effort was made to reproduce claims of low-frequency
(``LF'') pulses \cite{Agasa}, which for our purposes will be
taken to involve frequencies below 500~kHz.  (The AM broadcast
band contains numerous signals above 530~kHz, preventing the
study of higher frequencies.) Initial surveys were performed
using a Sony SW-7600G all-band portable receiver and an ICOM
IC-706 amateur transceiver.  However, considerably greater
sensitivity was achieved using a Palomar VLF converter which
converts the band 10--500~kHz to 3510--4000~kHZ, which was then
detected using the IC-706.

The major source of interference at the site was a nondirectional
aircraft beacon (NDB) operating at the Dugway airport on 284
~kHz.  Other NDBs and other LF carriers above about 110~kHz were
detectable but considerably weaker. A custom-made filter was
procured \cite{LFfil} to suppress the carrier at 284~kHz and
signals from the AM broadcast band above 500~kHz.  This filter was
employed during some of the low-frequency studies to be described
below.

\subsection{Measurement considerations and initial setup}

As mentioned above, the location of the receiving antenna (about
30~m east of the edge of the CASA array, at $x = 263.8$~m, $y =
0$~m), was dictated by a compromise between proximity to the
array and reduction of noise.  This noise was carried, to a large
extent, by the lightning-protection grid which overlays the array.

It was decided at an early stage to concentrate on the search for
horizontally polarized pulses as described in Section 2.
Consequently, a broad-band antenna with linear polarization was
adopted.  Initial surveys were taken with one of the original
antennas from the Mt.~Chacaltaya experiments \cite{Chac}, which
had been preserved from the 1960s.  This antenna was a large model
manufactured for VHF television reception, with some elements
which had been added by the experimenters to improve
low-frequency response.

The Mt.~Chacaltaya antenna was mounted on a portable searchlight
tower attached to a small trailer.  The tower could be extended
to a height of about 35 feet. The antenna was slightly damaged in
a collapse of the tower as a result of improper latching
procedures.  As insurance against further such incidents, a
portable military surplus log-periodic antenna was acquired. This
antenna (a Dorne and Margolin model to be described below)
was found to have superior response in the frequency
range of interest and very robust construction (even surviving a
subsequent tower collapse), and was adopted for subsequent studies.

The antenna was
mounted on the fully-extended searchlight tower with its center
at a height of 35 feet above ground, with the favored direction
of reception arriving from the zenith, and with arbitrary
azimuthal orientation.  Data were taken with two orientations:
``East-West'' polarization and ``North-South'' polarization, both
referred to magnetic North (14$^\circ$ east of true North
\cite{mag} at Dugway).  In addition, a projecting arm of the
mounting bracket was used to suspend a 10-meter-long vertical
antenna which was used for the low-frequency surveys.

The bandwidth to be covered by the RF search was not initially
specified, but to be determined by experience with survey
experiments.  Consequently, two main modes were used, a
narrow-band mode and a broad-band mode.  These are compared in
Table~2, where we also list a low-frequency mode used in the LF
survey. Their implementation is described in Sec.~\ref{ss:filt}.

\begin{table}
\caption{Modes of filtering.  (a) Suppression at 284~kHz and
above 500~kHz in some runs.}
\begin{center}
\begin{tabular}{c c} \hline
Mode & 3~dB bandpass (MHz) \\ \hline
Narrow-band & 25--35 \\
Broad-band & 25--250 \\
Low-frequency & 0.05--2.5 (a) \\ \hline
\end{tabular}
\end{center}
\end{table}

Some previous investigations (e.g., \cite{Chac} and \cite{Allan})
were able to detect RF pulses using a ``stand-alone'' trigger
based on the reception of transients alone.  This possibility was
investigated using a broad-band receiver with filters admitting
several different frequency ranges, and demanding coincidences of
signals received in a minimum number of channels.  It was found that
the vast majority of such transients at the Dugway site were not associated
with CASA/MIA events; they were probably due to atmospheric discharges. 
Such ``stand-alone'' transients, in fact, were found to increase during
periods of enhanced atmospheric electrical activity.  As a result, our main
results concern RF data taken with a trigger based on large CASA/MIA events.
We comment further on the possibility of a ``stand-alone'' trigger for
future experiments in Section~6.4.

The trigger was formed at the central CASA/MIA trailer, in a
manner to be described in detail below.  It was communicated to
the RF trailer over RG-59 cable.  The electrical length of the cable was 
found to correspond to a pulse delay of 2.15~$\mu$s.  No evidence for
pickup of this trigger pulse from the antenna was found.  Other
methods considered, and rejected in favor of the simpler
electrical communication, included optical fiber and infrared
sensors.

\subsection{Design features}

\subsubsection{Antenna system}

A portable log-periodic antenna manufactured by Dorne and Margolin, Model
\# DM ARM 160-5, with a nominal response of 30--76~MHz, was acquired from
FairRadio Co.~in Lima, Ohio, for about
\$60.  (A spare was used for noise studies at the University of
Washington.) Overload protection was provided by two 1N4148
diodes of opposite polarity connected across the antenna
terminals, leading to a maximum output voltage of about $\pm 0.6$~
V.  A gas discharge tube manufactured by Alpha/Delta provided
lightning protection.  Some properties of the antenna are
summarized in Table 3.

\begin{table}
\caption{Properties of log-periodic antenna used for RF studies}
\begin{center}
\begin{tabular}{l c} \hline
Nominal frequency range (MHz) & 30--76 \\
Usable frequency range (MHz) & 28--170 \\
Number of elements & 9 \\
Dimensions (m) & $3 \times 3$ \\
Feedline RG-58U & 60 feet \\ \hline
\end{tabular}
\end{center}
\end{table}

\subsubsection{RF front-end}

The RF amplification stage consisted primarily of one or two
ZFL-500LN low-noise broad-band preamplifiers manufactured by
Mini-Circuits, and for certain runs low-noise preamplifiers
manufactured by ANZAC.  Specifications of these preamplifiers are
summarized in Table 4.

\begin{table}
\caption{Properties of preamplifiers used for RF studies}
\begin{center}
\begin{tabular}{l c c c c} \hline
Manufacturer &   Model    & DC power & Gain & Frequency \\
             &            &   (V)    & (dB) & range (MHz) \\ \hline
Mini-Circuits & ZFL-500LN &  13.6    &  26  &  DC--500  \\
ANZAC        &   AM-107   &   18     &  10  &   1--500  \\ \hline
\end{tabular}
\end{center}
\end{table}

\subsubsection{Filtering} \label{ss:filt}

Table 5 contains a summary of all filters used in the experiment
with the exception of the 284~kHz filter \cite{LFfil} described
previously. These filters are manufactured by Mini-Circuits; they
were obtained with tubular cases fitted with BNC connectors.

\begin{table}
\caption{Filters used in RF data acquisition}
\begin{center}
\begin{tabular}{l c c} \hline
Model &    Type     & 3~dB point(s) \\
      &             & (MHz) \\ \hline
BLP-1.9 & Low-pass  & 2.5 \\
BHP-25  & High-pass & 25 \\
BLP-30  & Low-pass  & 35 \\
BBP-30  & Bandpass  & 25, 35 \\
BLP-250 & Low-pass  & 250 \\ \hline
\end{tabular}
\end{center}
\end{table}

A typical ``narrow-band'' configuration described in Table~2
involved feeding the signal from the antenna through the feedline,
a BHP-25 filter and a BLP-30 filter with combined 3~dB points of 25 and 35
~MHz, a ZFL-500LN preamplifier with 26~dB of gain, a BBP-30 filter
with 3~dB points 25 and 35~MHz, another Mini-Circuits ZFL-500LN
preamplifier with 26~dB of gain, and a BHP-250 filter to suppress
any high-frequency noise.  (Some data runs involved permutations
of these components.  The above configuration was  found to
minimize feed-through of preamplifier noise.  Some runs involved
a dual ANZAC preamplifier instead of a ZFL-500LN.)

A ``broad-band'' configuration involved the same feedline and
BHP-25 filter, a single ZFL-500LN preamplifier, and a BLP-250 filter.  A
``low-frequency'' configuration involved the feedline,
a BLP-1.9 filter and a
ZFL-500LN preamplifier, with a 284~kHz notch filter inserted
before or after the BLP-1.9 in some runs.  The notch filter also
contained a roll-off above 500~kHz.

\subsubsection{``Large-event'' trigger and design}

A trigger based on the coincidence of seven of the eight outer
muon ``patches'' (see Fig.~1) was set to select ``large'' showers
in the following manner. Each muon patch was set to produce a
trigger pulse of length 5~$\mu$s and amplitude $-120$~mV when $n$
of its 64 counters registered a minimum-ionizing pulse within $5.2$
~$\mu$s of one another.  For engineering runs (until 12/28/96),
$n$ was set equal to 4, while for later runs it was increased to
5 to favor larger showers and reduce noise.  The pulses
were then combined in two groups of 4, feeding through two 2X
attenuators into two fan-in/fan-outs (to avoid saturation of
inputs) and the resulting pulses further combined to produce a
summed pulse.  This signal was fed to a LeCroy 821 Discriminator,
whose output was amplified to an amplitude of about $-6$~V and
then sent over RG-59 cable to the RF trailer (see Fig.\ 1).  The
trigger pulse at the RF trailer had an amplitude of about $-2.4$
~V and a duration of $1~\mu$s.

The above trigger was estimated to correspond to a minimum shower
energy of somewhat below $10^{16}$~eV, based on the integral rate
\cite{REF} at $10^{18}$~eV of 0.17/km$^2$/day/sr. At this level
good correlation could be established between trigger pulses and
events recorded by the CASA data acquisition system. 

Only shower radiation that is stronger than 3~$\mu$V$/$m$/$MHz can exceed the
average noise level at CASA site by three standard deviations or more
[Sec.~4.6]. According to the original Haverah Park results [Eq.~(\ref{eqn:E})],
a typical shower that would lead to such radiation is a vertical shower of
energy $10^{17}$~eV or higher at a distance of 210~m. If the rate for showers
with energy greater than $E$ behaves as $1/E^2$, showers above $10^{17}$~eV
would be expected to occur with a rate of 17/km$^2$/day/sr.  Since RF detection
relies on muon triggering, the antenna can only detect radiation from those
showers whose cores pass inside the rectangle of the array. Just about
0.06~km$^2$ of the array area lies inside the 210~m radius from the antenna.
With the solid angle of observation limited by a zenith angle of 50$^\circ$,
one expects about 2.25 detectable shower pulses per day.

\subsubsection{Data acquisition}

A Tektronix TDS-540B digitizing oscilloscope registered filtered
and preamplified RF data on a rolling basis. These data were
captured and stored on hard disk using a National Instruments
GPIB interface upon receipt of a large-event trigger.  Data were
taken using various computers at different times, allowing
for analysis both at the University of Washington and at
Chicago.  The Washington system used a Macintosh Quadra 950
running Labview, with a latency time of about 8 seconds between
events, while the Chicago system used either a Dell XPS200s desktop
or a Dell Latitude LM laptop running a C program adapted from those
provided by National Instruments, with a latency time of about 2 seconds.
Each trigger caused 50~$\mu$s of RF data, centered around the
trigger and acquired at 1~GSa/s, to be saved.

\subsubsection{Rates and off-line processing}

The total trigger rate ranged between about 20 and 50 events per
hour, depending on the value of $n = 4$ or $5$ of muon counters
chosen to generate a patch trigger pulse and on intermittent
sources of noise sometimes present in the trigger system.
Concurrently, the CASA on-line data acquisition system was
instructed via a special program called MUTRIG to write files of
events in which at least 7 out of the 8 outermost muon patches
produced a patch pulse.  These files, one for each CASA run,
typically overlapped with the records taken at the RF trailer to
a good but not perfect extent \cite{mutrig} as a result of occasional noise 
on
the trigger line.  Moreover, an undiagnosed timing problem
occasionally caused the loss of a muon trigger pulse for certain
large events recorded by MUTRIG.

\subsubsection{Off-line overload rejection}

Events were typically recorded at a gain such that the maxima and
minima corresponded to about 2/3 of the dynamic range of the
oscilloscope's 8-bit data acquisition system (ranging from $-128$
to $127$ digitization units). Local intermittent monochromatic RF
signals occasionally saturated this dynamic range.  Such events
were rejected off-line by discarding any cases in which the
maxima and minima exceeded 100 digitization units.
In the configuration used for the final bounds on pulse amplitudes,
corresponding to an oscilloscope setting of 5~mV per division of 25
digitization units, we thus rejected all signals corresponding to
preamplifier peak outputs greater than $\pm 20$~mV.  In some cases, with
oscilloscope settings of 20~mV per division, we rejected signals with
preamplifier peak outputs greater than $\pm 80$~mV.  In all cases these
voltages were well below the manufacturer's specified limit of
$\pm 400 \sqrt{2}$~mV (3~dBm), and within a satisfactorily linear range of
preamplifier response.

\subsubsection{Calibration}

The average gain $G_{ant}$ of the antenna in its forward direction rises 
from about 3~dBi (decibels with respect to an isotropic radiator)
at 30~MHz to a peak of about 5~dBi at 50~MHz, slowly decreasing
to 4~dBi at 76~MHz \cite{PM}.  We shall take an average
gain of 4~dBi ($G_{ant}=2.5$) over the frequency range of interest.

More precise calibration would involve modelling of the gain
pattern using a program such as EZNEC \cite{EZNEC}, and integrating
over directions of expected signal arrival.
This modelling also would have been useful in order to emulate the
frequency-dependent phase distortion induced by the antenna, but it was
not found possible to obtain a sufficiently close fit to the antenna's
measured standing-wave-ratio characteristics to take such a model
seriously.  Certainly this point should be addressed in any future
studies.  One might also utilize
sources of known strength such as amateur radio satellites
broadcasting on 29.4~MHz, FM and television stations, and galactic and
solar noise.  The existing data contain signals from FM and television
stations broadcasting near Dugway, some of whose field strengths are well
enough known that they may be usable for calibration.  Alternatively, for
future work it would be helpful to calibrate antennas on an antenna range
at some distance from an impulse generator, broadcasting through a
broad-band antenna with already-determined characteristics.

\subsection{Signal processing}

\subsubsection{Fourier methods} \label{ss:Four}

In order to remove strong Fourier components associated with
signals which were approximately constant over the duration of
each data record, a short MATLAB routine was written to perform
the fast Fourier transform of the signal and renormalize the
large Fourier components to a given maximum intensity.  Fig.~2
shows the fast Fourier transform of a typical RF signal before
and after this procedure was applied. In each case the data were
acquired using the ``wide-band'' filter configuration, whose
response cuts off sharply below 23~MHz.

\begin{figure}
\centerline{\epsfysize = 5 in \epsffile{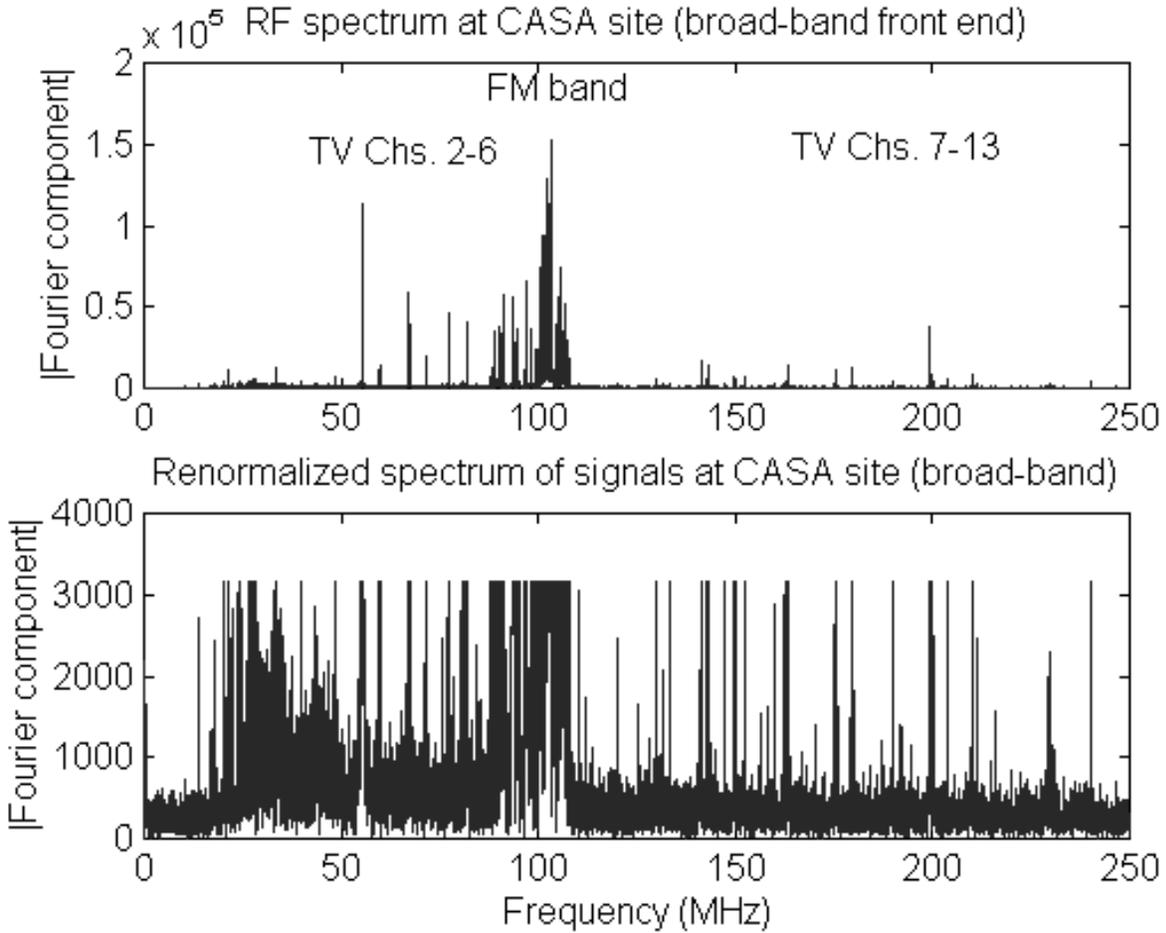}} \caption{Top
panel: Fourier spectrum (in arbitrary units) of RF signals
acquired at Dugway site using high-pass 25~MHz and low-pass 
250~MHz filters. Prominent features include video and audio carriers
for TV Channels 2, 4, 5, 7, and 11 (see Table 1 for frequencies),
and the FM broadcast band between 88 and 108~MHz. Bottom panel:
Fourier spectrum (same vertical scale) after renormalization of
large Fourier components to a magnitude chosen here to be $3.16
\times 10^3$.  In practice best sensitivity to transients was
obtained by renormalizing to a magnitude of $10^3$.}
\end{figure}

The effect of digital filtering on detectability of a transient is
illustrated in Fig.~3.  The top panel shows the RF record whose
Fourier transform was given in Fig.~2, on which has been
superposed a simulated transient of peak amplitude 14.5
digitization units. (The data acquisition scale ranges from $-128$
to $+127$ digitization units; one scale division on the
oscilloscope corresponds to 25 units.)  The transient is invisible
beneath the large amplitude associated with television and FM
radio signals.  The middle panel shows the result after
application of the Fourier coefficient shrinkage algorithm.

\begin{figure}
\centerline{\epsfysize = 7.5 in \epsffile{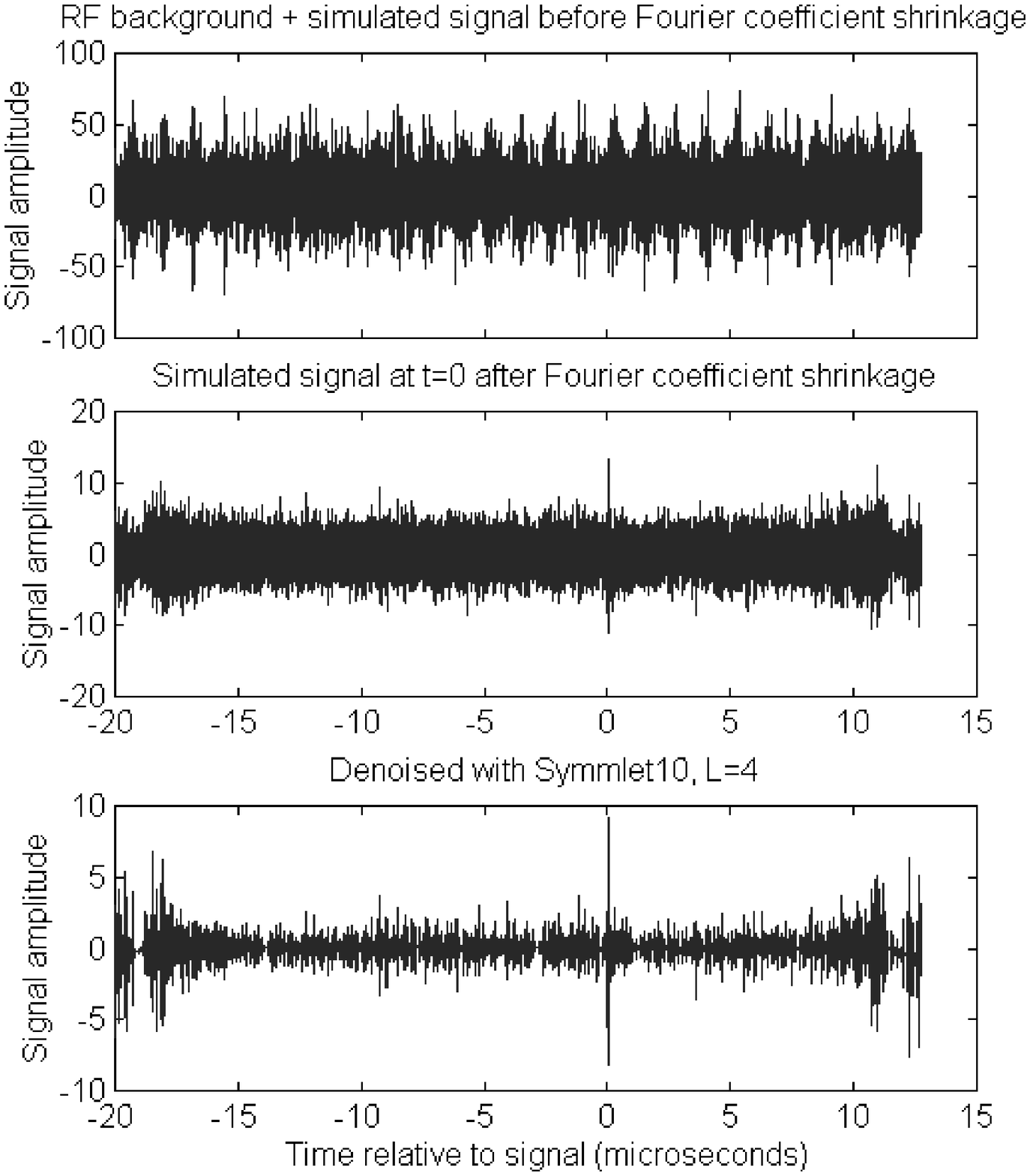}}
\caption{Effect of Fourier coefficient shrinkage on detectability
of a transient.  Top panel:  raw RF record (in arbitrary units)
with simulated signal superposed.  Middle panel:  record (same
scale) after Fourier coefficient shrinkage.  Here a maximum
Fourier coefficient magnitude of $10^3$ (in the units of Fig.~2)
has been imposed.  Bottom panel: the same record after denoising
with a 10-point symmlet level $L=4$ routine \cite{DFS}.}
\end{figure}

The event in Figs.~2 and 3 consisted of 32,768 data points
obtained at a 1~ns sampling interval, with the trigger at the
20,000th point.  The frequency resolution in the fast Fourier
transform is thus 500~MHz (the Nyquist frequency) divided by
16,384, or about 30~kHz.  This permits rather fine distinction
between frequencies containing a strong carrier and those which
correspond to its weaker sidebands.  At the same time, it permits
time resolution to be preserved, allowing for the examination of
rather rapid transients.  To the extent that these transients do
not contain Fourier coefficients exceeding a pre-determined
threshold, they should be relatively unaffected by the shrinkage
algorithm in the absence of interfering signals. However, since
at Dugway signals in nearly the whole FM band (88--108~MHz)
exceed the threshold, some distortion is unavoidable
using such a method.  In obtaining bounds on pulse amplitudes we therefore
employ a method involving the comparison of Fourier power in a given
time window with the average power obtained over the whole data record
for each Fourier component.  This method is described below.

\subsubsection{Time-frequency analyses} \label{ss:tfa}

One can perform a fast Fourier transform using a small time
window (typically 1024~ns) which is advanced sequentially through
the data record, typically in steps of 100~ns.  The frequency
resolution of any given ``snapshot'' is then 500~MHz divided by
(typically) 512, or a bit better than 1~MHz.  A two-dimensional
display of time vs.~frequency then allows one to distinguish
short transients (with components over many frequency bins) from
continuous RF sources (with components in narrow frequency ranges
over the entire time record).  One such plot appears in Fig.~11,
Sec.~5.3.1, below. In practice one may wish to suppress
frequencies corresponding to the whole FM band and known TV
stations, so as not to overload the dynamic range of the
display.  An alternative method \cite{Gross} is to renormalize
each point in time--frequency space so that {\it deviations} from
the average in each frequency bin are displayed.  This method is
described further in Sec.~5.2. 
It was used for the main part of data analysis.

\subsubsection{Wavelet techniques}

The wavelet package {\tt Wavelab} \cite{WL} contains a denoising
routine which was adapted for our purposes.  While an exhaustive
search for optimized methods was not performed, good results in
reducing noise levels were obtained using a 10-point symmlet
routine with level $L=4$ \cite{DFS}.  An example of a denoised
signal is shown in the bottom panel of Fig.~3.  Here a simulated
signal of positive peak amplitude 14.5 digitization units has been
added to an RF record otherwise free of transients. The effect of
wavelet denoising is to reduce the amplitude of random
high-frequency fluctuations while preserving edge effects such as
transients.

\subsection{Signal simulation}

We wished to quantify the improvement associated with each method
of signal processing.  We thus simulated the expected signal by
generating it using an arbitrary waveform generator, feeding it
through the same preamplifier and filter configurations used for
data acquisition, and superposing it on records otherwise free of
transients.  We successively reduced the amplitude of the
superposed test signal until it could not be distinguished from
random noise peaks, thereby obtaining an estimate of sensitivity.

A Hewlett-Packard Arbitrary Waveform Generator was used to
generate signals whose characteristics are illustrated in Fig.~4.
These signals were taken to have the form $f(t) = \theta(t)
At^2(e^{-Bt} - C e^{-Dt})$ with the coefficient $C$ chosen so
that $f(t)$ has no DC component, and $D$ corresponding to a long
duration of the negative-amplitude component.  For all pulses we
chose $D = B/20$, so that $C = (8000)^{-1}$ cancels the DC
component.  The Fourier components of the test pulse fall off
smoothly with frequency.  The initial $t^2$ behavior was chosen
so that both the test pulse and its first derivative vanish at
$t=0$, as might be expected for a pulse from a developing shower.

The simulated pulses are summarized in Table 6.  Instead of
quoting the value of $A$, we quote the maximum positive value of
the pulse, both before and after filtration and preamplification.
These values of $V_{\rm pk}$ reflect choices for convenience in display on
the oscilloscope, and are otherwise arbitrary.

\begin{figure}
\centerline{\epsfysize = 5 in \epsffile{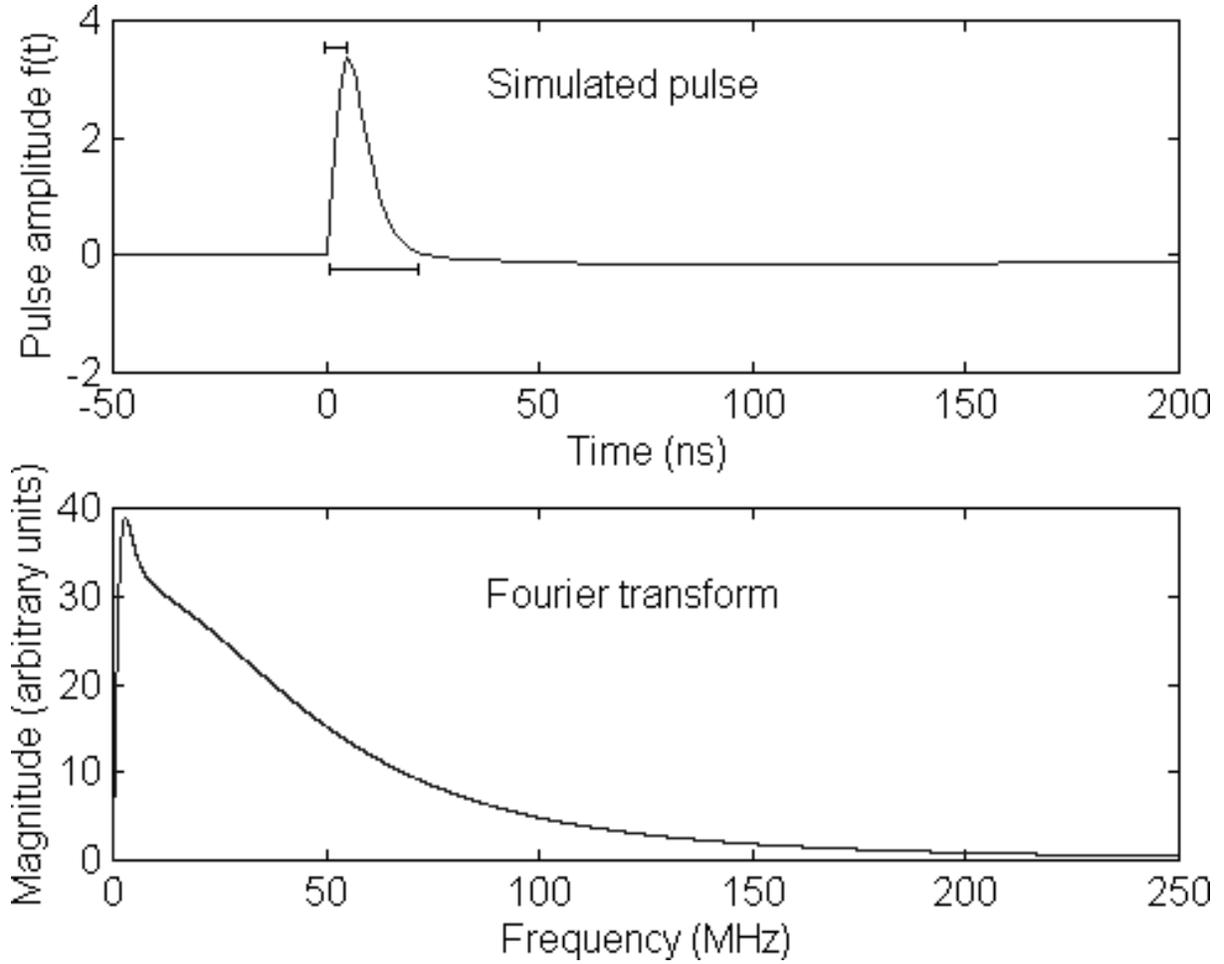}}
\caption{Analytic depiction of typical pulse presented to
filter-preamplifier configuration.  Top panel:  time dependence
of pulse $f(t) = \theta(t) t^2[e^{-0.4t} - e^{-0.02t}/ 8000] (t$
in ns); bottom panel: Fourier spectrum of pulse 
(calculated analytically). In the top
panel, the short bar above the pulse denotes $\delta$, the time difference
between onset and maximum, while the longer bar below the pulse denotes
$\Delta$, the duration of the positive component.}
\end{figure}

\begin{table}
\caption{Parameters of test signals.  $\delta$ is the time
between pulse onset and maximum, while $\Delta$ is the duration
of the positive component of the pulse.  $V_{\rm pk}$ is the peak
(positive) input voltage to the filter-preamplifier
configuration.  The letter after the peak voltage denotes (a)
narrow-band (25-35~MHz) or (b) broad-band ($> 25$~MHz)
configuration (see Sec.~\ref{ss:filt}).  $V_{\rm out}$ is the
peak-to-peak amplitude of the pulse emerging from the
filter-preamplifier configuration. $S$ is the scale factor with
which data were recorded on oscilloscope.}
\begin{center}
\begin{tabular}{c c c c c c} \hline
$B$ & $\delta$ & $\Delta$ & $V_{\rm pk}$ & $V_{\rm out}$ & S  \\
(ns$^{-1}$) &   (ns)   &    (ns)  &  (mV)   & (mV) & (mV/div) \\
\hline
   0.8      &   2.5    &    12    &  1.2 (a) &  86  &   20    \\
            &          &          &  6.0 (b) & 124  &   20    \\
   0.4      &    5     &    24    &  0.7 (a) &  70  &   20    \\
            &          &          &  1.3 (b) &  21  &    5    \\
   0.2      &   10     &    47    &  0.7 (a) &  71  &   10    \\
            &          &          &  7.0 (b) &  67  &   10    \\
   0.1      &   20     &    95    &  1.5 (a) &  75  &   10    \\
            &          &          &  7.6 (b) &  32  &    5    \\ \hline
\end{tabular}
\end{center}
\end{table}

The shape of the pulse of Fig.~4 is affected by preamplification
and filtration as shown in Figs.~5 (broad-band) and 6
(narrow-band). The noise in these figures and the sharp feature
at 125~MHz in Fig.~5 are associated with the system used to
generate the test pulse, and the fact that the Fourier transform
is taken over a much longer time than the duration of the pulse.

Systematic studies of signal-to-noise ratios have been performed
so far only for the simulated pulses with $\delta=5$~ns applied
to a broad-band front end [(b) in Table 6].  The value of
$\delta$ is a measure of the distance $R$ of closest approach of
the shower core \cite{Allan}.  This choice corresponds to a
typical distance $R \simeq 200$~m.  A typical pulse of this type
gave a front end output of 21~mV peak-to-peak, acquired at an
oscilloscope sensitivity of 5~mV per division.  Each division
corresponds to 25 digitization units, so the peak-to-peak range
is about 104 digitization units, or slightly less than half the
dynamic range (255 units, or 8 bits).  Positive and negative
peaks are thus about 52 digitization units each.

The stored test signal is then multiplied by a scale factor and
added algebraically to a collection of RF records in which, in
general, randomly occurring transients will be present.  One then
inspects these records to see if the transient can be
distinguished from random noise.

\begin{figure}
\centerline{\epsfysize = 5 in \epsffile{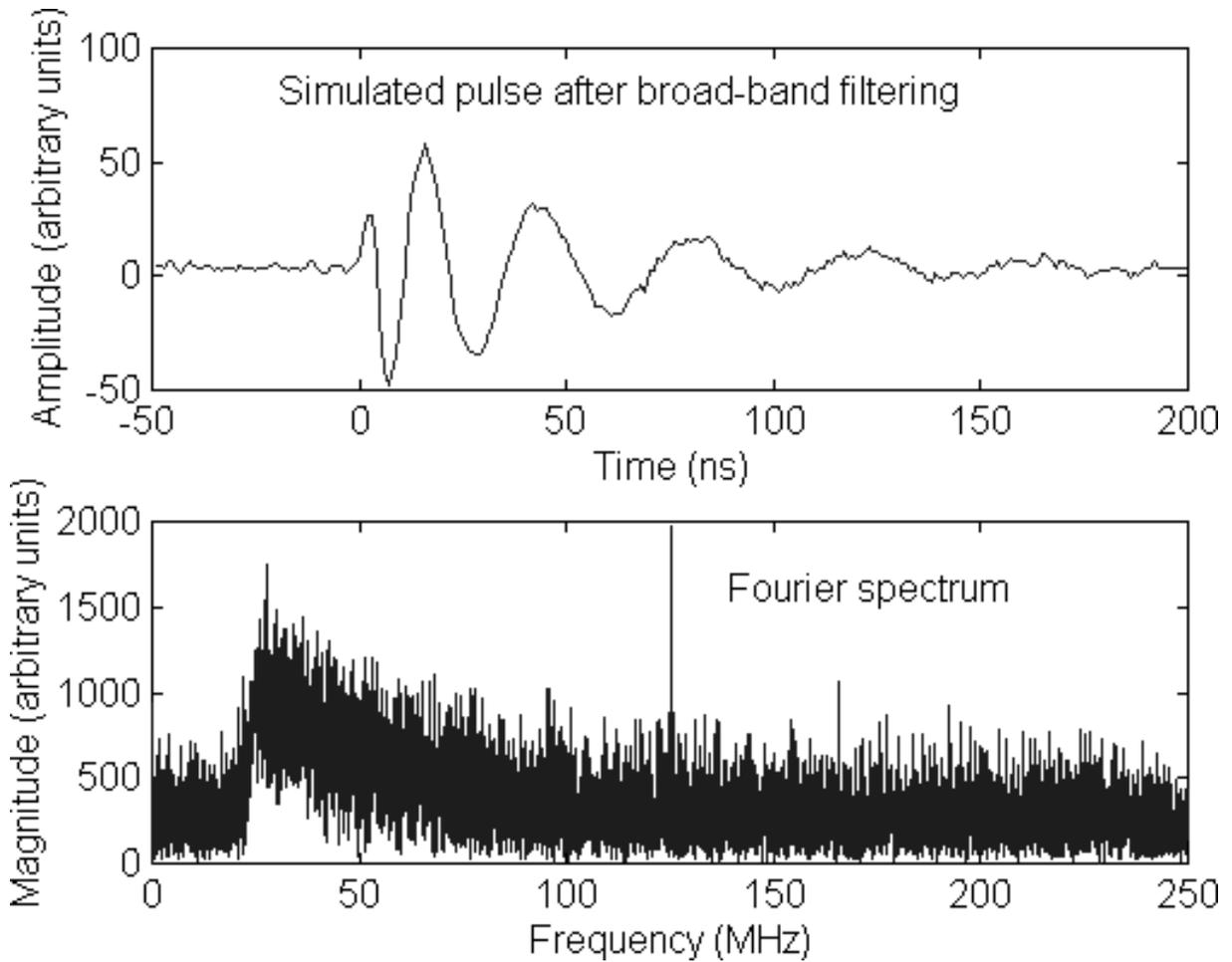}} \caption{Test
pulse of Fig.~4 after broad-band filtration ($> 25$~MHz) and
preamplification. Top panel:  time dependence of pulse; bottom
panel: Fourier spectrum of recorded pulse for $-20~\mu {\rm s}
\le t \le 12.768~\mu$s.}
\end{figure}

For the broad-band data we estimated that pulses with input
voltages corresponding to about 1/5 the original test pulse amplitude can
be distinguished from average noise (not from noise spikes!).
Since the original test pulse had a peak value of 1.3~mV, this
corresponds to sensitivity to an antenna output of about $V_{\rm
pk} \simeq 260~\mu$V.  The ability to detect such a pulse with an
effective bandwidth of about 30~MHz corresponds to a threshold
sensitivity at the level of order $3~\mu$V/m/MHz (see Appendix B).

Preliminary studies of simulated pulses applied to the
narrow-band front end suggest a considerably poorer achievable
signal-to-noise ratio, despite the expectation that the signal
should have a large portion of its energy between 23 and 37~MHz.
It appears difficult to detect a pulse from the antenna below
about 0.7~mV, which for a bandwidth of 14~MHz corresponds to a
threshold sensitivity of $7~\mu$V/m/MHz, not sufficient for our
purposes. Studies of possible improvements of the analysis
algorithm for the narrow-band data are continuing.

\begin{figure}
\centerline{\epsfysize = 5 in \epsffile{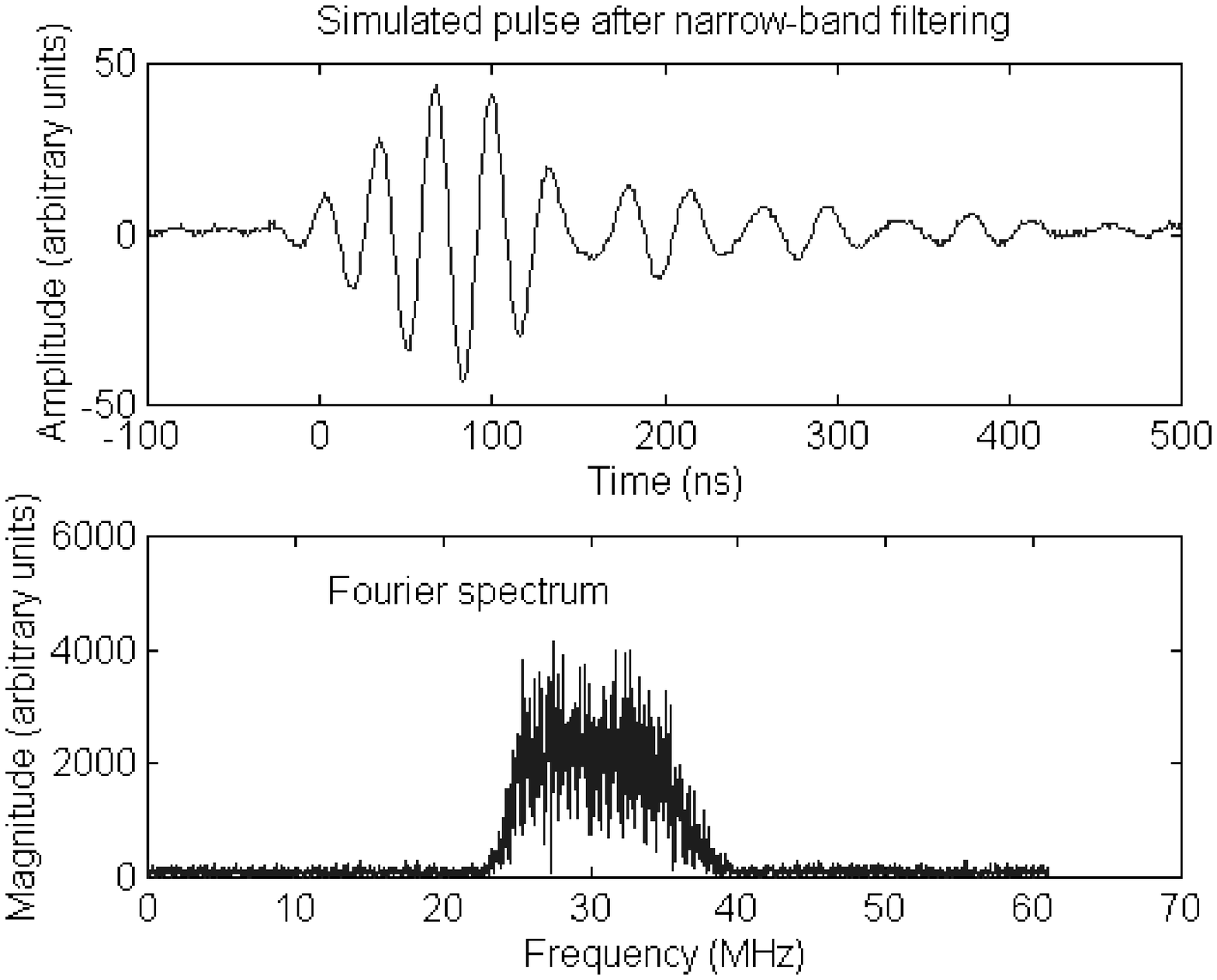}} \caption{Test
pulse of Fig.~4 after narrow-band filtration (25--35~MHz) and
preamplification.  Top panel: time dependence; bottom panel:
Fourier spectrum of recorded pulse for $-20~\mu {\rm s} \le t \le
12.768~\mu$s.}
\end{figure}

\section{Results}

\begin{table}
\caption{Triggers associated with CASA operation taken under
various conditions.}
\begin{center}
\begin{tabular}{l r r r} \hline
Front end   & Macintosh         & Dell & Total \\ \hline
Narrow-band &  5849 (139.78 h) & 1952 (48.97 h) & 7801 (188.75 h) \\
Broad-band  &  9603 (272.57 h) & 5416 (121.62 h) & 15019 (394.18 h) \\
Low-frequency &    0            & 505  (17.67 h) & 505 (17.67 h) \\ \hline
Total       & 15452 (412.35 h) & 7873 (188.25 h) & 23325 (600.6 h) \\
\end{tabular}
\end{center}
\end{table}

\begin{table}
\caption{Broad-band data recorded on Macintosh Quadra.}
\begin{center}
\begin{tabular}{l c c c c} \hline
Antenna      & CASA           &  CASA            & Partial & Total  \\
Polarization & HV on          & HV off           & CASA HV & events \\ \hline
East-West    & 4966 (119.03 h) & 859 (21.88 h) & 1957 (53.08 h) & 7782 
(194.0 h)  \\
North-South  & 696 (30.53 h)   & 582 (23.25 h) & 543 (24.78 h) & 1821 
(78.57 h) \\ \hline 
Total events & 5662 (149.57 h) & 1441 (45.14 h) & 2500 (77.87 h)
         &  9603 (272.57 h)  \\ \hline
\end{tabular}
\end{center}
\end{table} 

\subsection{Event sample}

More than 20000 triggers, obtained under various conditions of
filtering, preamplification, 
and noise reduction during the period February 1997 -- March 1998,
are summarized in Table 7.  Events recorded on a Macintosh
Quadra 950 and those recorded on a Dell LM Latitude laptop
computer are listed separately because the power supply of the
latter introduced spurious transients.  Our initial analysis concentrated on 
data taken with
the Macintosh.  For reasons mentioned above, we consider only the
broad-band data at this time.  Thus, our usable sample consists of
over 9000 CASA triggers.  In addition, periodic forced
triggers were taken to monitor noise activity not associated with
CASA operation.

The broad-band data recorded on the Macintosh Quadra, summarized in Table~8,
are subdivided into several categories.  Data were taken with both
East-West (EW) and North-South (NS) antenna polarizations. Moreover, since
noise from CASA boxes was found to be a significant source of RF
transients, data were taken with some or all CASA boxes disabled
by turning off the high voltage (HV) supply to the photomultipliers. 
Even when HV is supplied only to boxes that are further than 100~m from the 
antenna, the RF transients from these boxes provide a strong background.  (See 
the discussion in Sec.~\ref{s:Char} and Fig.~10 below.)
One is unlikely to distinguish the RF pulses of the showers from this 
noise. Therefore, we concentrated on data with CASA HV off,
with 859 triggers taken with EW antenna polarization (21.88 active hours) 
and 582 triggers taken with NS antenna polarization (23.25 active hours). 
For subsequent sensitivity calculations, a subset of data was
used consisting of 756 EW triggers (17.25 hours) and 528 NS triggers
(21.12 hours). The remaining data with CASA HV off occurred in very
short runs (52 EW triggers and 19 NS triggers) or was contaminated
by local VHF communication signals (51 EW triggers and 35 NS triggers).
Data with CASA HV on or partially disabled were not used for the
present analysis.

\begin{figure}
\centerline{\epsfysize = 5 in \epsffile{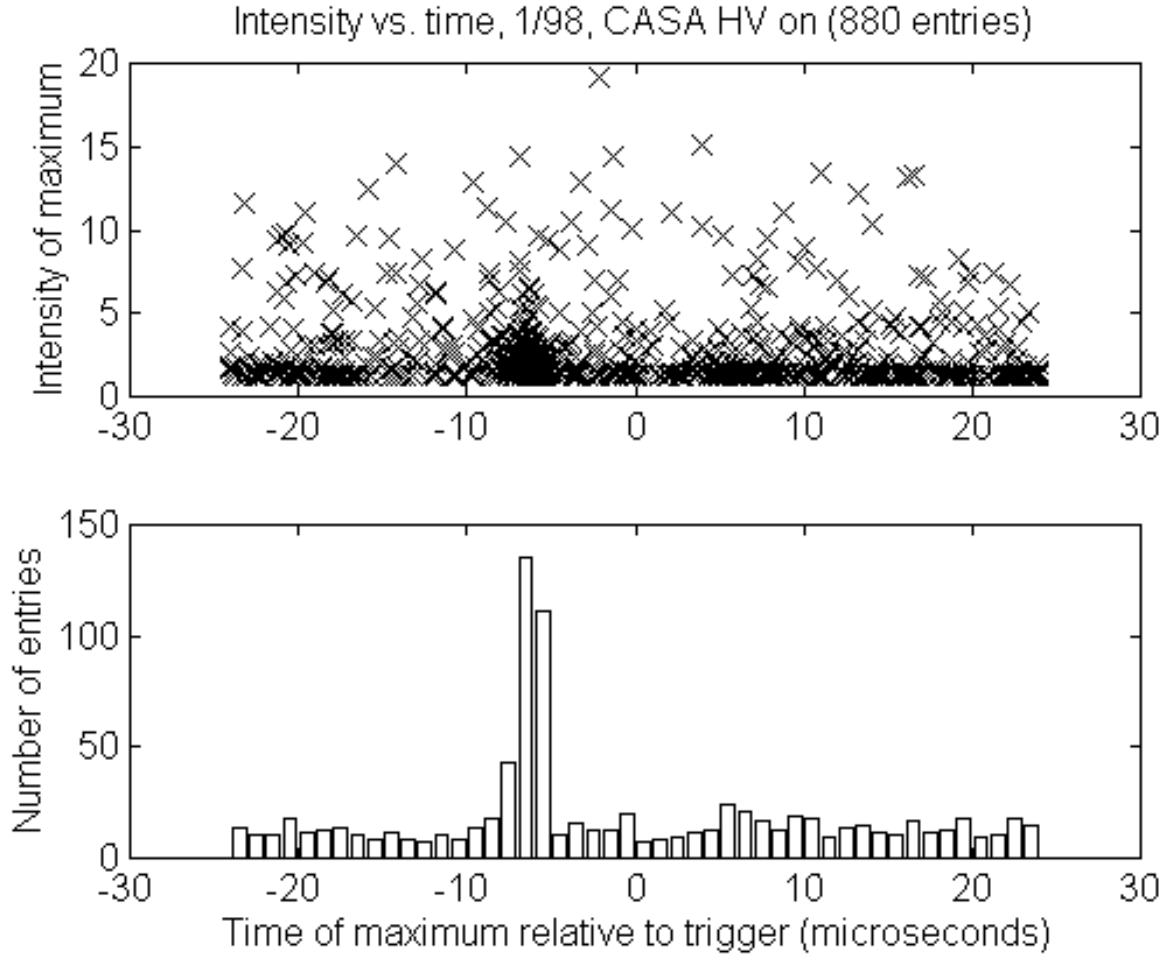}} \caption{Top
panel:  intensity-vs.-time plot for maxima of 880~pulses recorded in 
698~triggers in January 1998 with CASA HV supplied to all stations.
Bottom panel: time distribution of transients.
All events recorded with East-West antenna polarization.}
\end{figure}

\subsection{Characterization of transients associated with CASA operation}
\label{s:Char}

Several means were used to characterize transients.  One method with good 
time
resolution involved the shrinkage of large Fourier coefficients to a fixed
maximum intensity, as in Figs.~2 and 3.  Another, which we have used for
results to be presented below, involves generation of a time-vs.-frequency
intensity plot by Fourier-transforming 1024-ns subsets of the 50 $\mu$s data
record, spaced by 100 ns steps.  Since the data are sampled at 1 ns 
intervals,
the frequency resolution of this method is thus about 1 MHz.  The intensity
$S(\nu,t)$ is then averaged over time $t$ for each frequency $\nu$ to form
an average intensity $\bar S(\nu)$.  The quantity $S(\nu,t)/\bar S(\nu)$ is
an estimate of the degree to which the intensity at a given frequency $\nu$
and time $t$ exceeds the average over the $50~\mu$s sampling time.  We then
average $S(\nu,t)/\bar S(\nu)$ over $\nu$ to search for events in which the
average intensity at a given time is exceeded in many simultaneous frequency
bands.

\begin{figure}
\centerline{\epsfysize = 5.5 in \epsffile{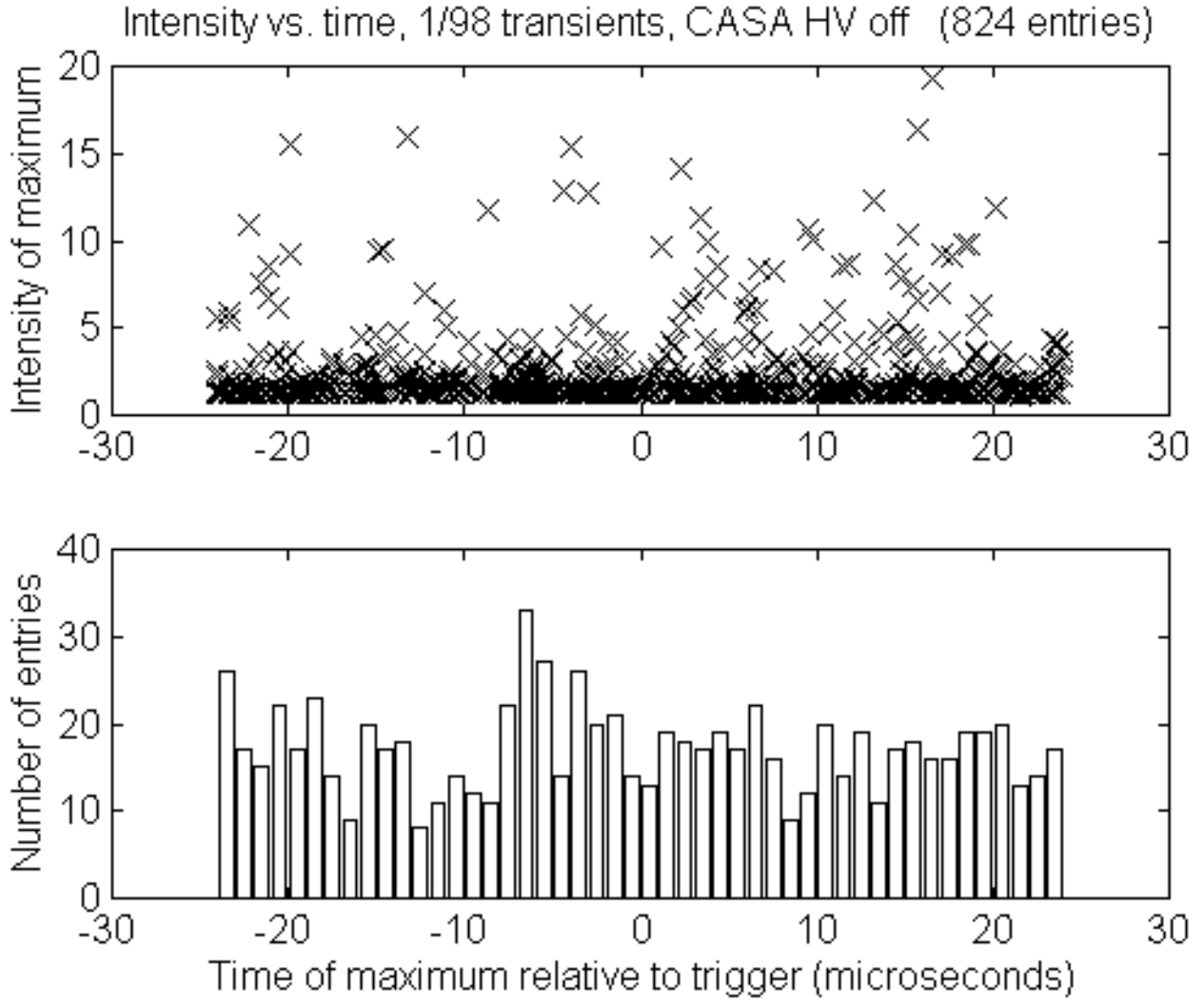}} \caption{Top
panel:  intensity-vs.-time plot for maxima of 824 pulses in 691 
triggers recorded
in January 1998 with CASA HV disabled.
Bottom panel:  time distribution of transients.
All events recorded with East-West antenna polarization.}
\end{figure}

One can then search for peaks of each data record 
(there may be several peaks in a record), 
plotting intensity of their maxima against time relative to the trigger. 
One such plot is shown in Fig.~7 for a data run in which CASA HV was
delivered to all boxes.  A strong accumulation of transients,
mostly with intensity just above the arbitrarily chosen threshold (mean + 3
$\sigma$ for each trigger sample), is visible at times $-5$ to $-7~\mu$s
relative to the trigger.  In a comparable plot for a run in which CASA HV was
completely disabled (Fig.~8), only a small accumulation at times $-6$ to
$-7~\mu$s is present.  This excess appears due to transients with
predominantly high-frequency components (over 100~MHz).  Since signal
pulses are expected to have more power below 100~MHz (see Fig.~5, bottom)
we believe that this accumulation is not due to shower radiation, but
most likely arises from the muon patches, one of which is within
75~m of the antenna.

\begin{figure}
\centerline{\epsfysize = 5 in \epsffile{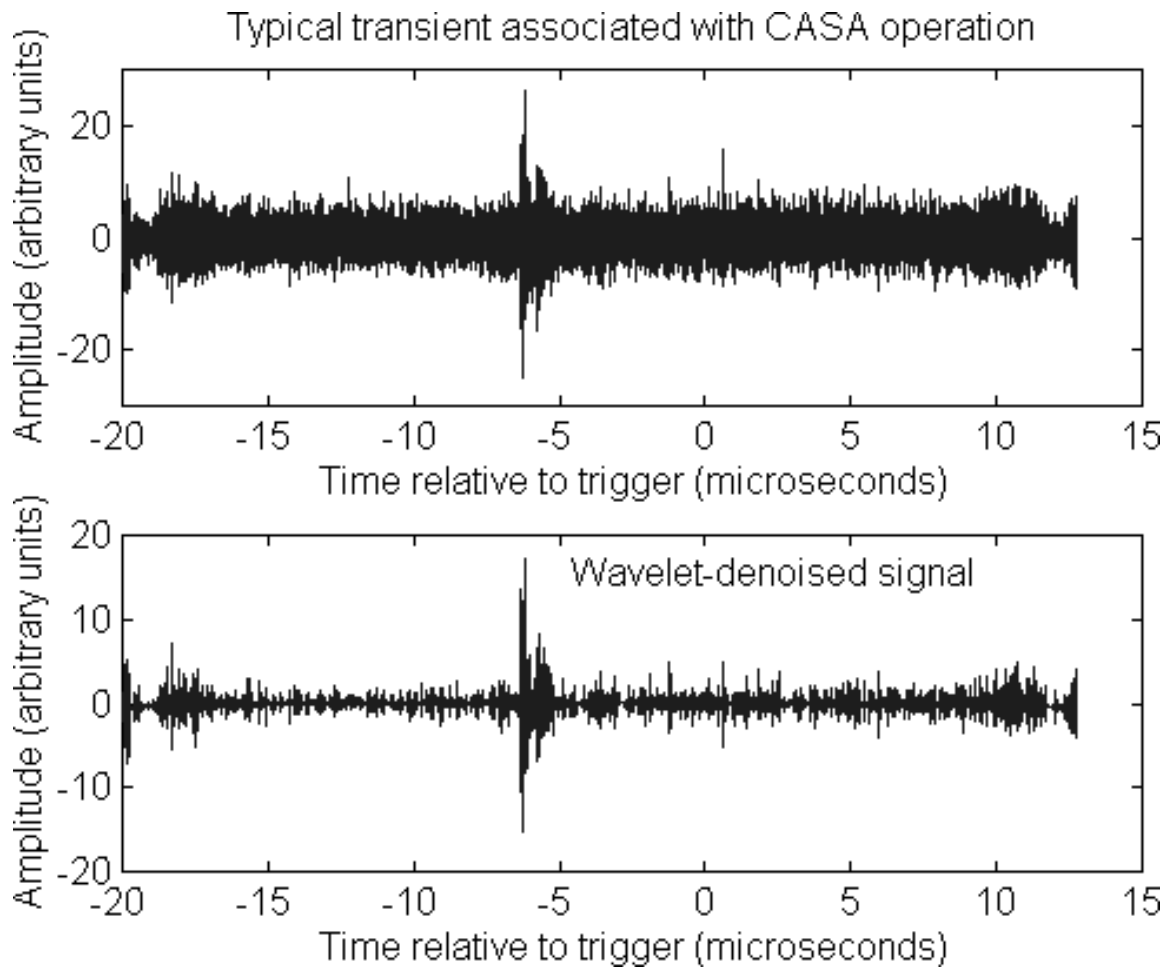}}
\caption{Signal of a typical transient associated with CASA
operation.  Top panel:  before denoising; bottom panel:  after
denoising.}
\end{figure}

A typical transient occurring in a run with CASA HV on is shown in Fig.~9.  
The transients are highly suppressed (though not in all runs) when
CASA boxes within 100~m of the antenna are disabled, as shown in
Fig.~10.

The time distribution of pulse maxima above an arbitrary
threshold for 880 pulses detected with CASA HV on (one run
from January 1998 composed of 698 files of data) is shown in 
the bottom panel of Fig.~7.  The
mean arrival time is about $6~\mu$s before the trigger, with a
distribution which is slightly broader for pulses arriving
earlier than the mean.  This broadening may correspond to some
jitter in forming the trigger pulse from the sum of muon patch
pulses.

As mentioned earlier, the time for the trigger pulse to propagate
from the central station to the RF trailer was measured to be
$2.15~\mu$s.  One expects a similar or slightly greater
travel time for pulses to
arrive from muon patches to the central station (see Fig.~1).
Moreover, the muon patch signals are subjected to delays so that
they all arrive at the central station at the same time for a
vertically incident shower.  Thus, the peak in Fig.~7 is
consistent with being associated with the initial detection of a
shower by CASA boxes.  This circumstance was checked by recording
CASA trigger request signals simultaneously with other data; they
coincide with transients such as those illustrated in Fig.~9
within better than $1/2~\mu$s.

\begin{figure}
\centerline{\epsfysize = 4.5 in \epsffile{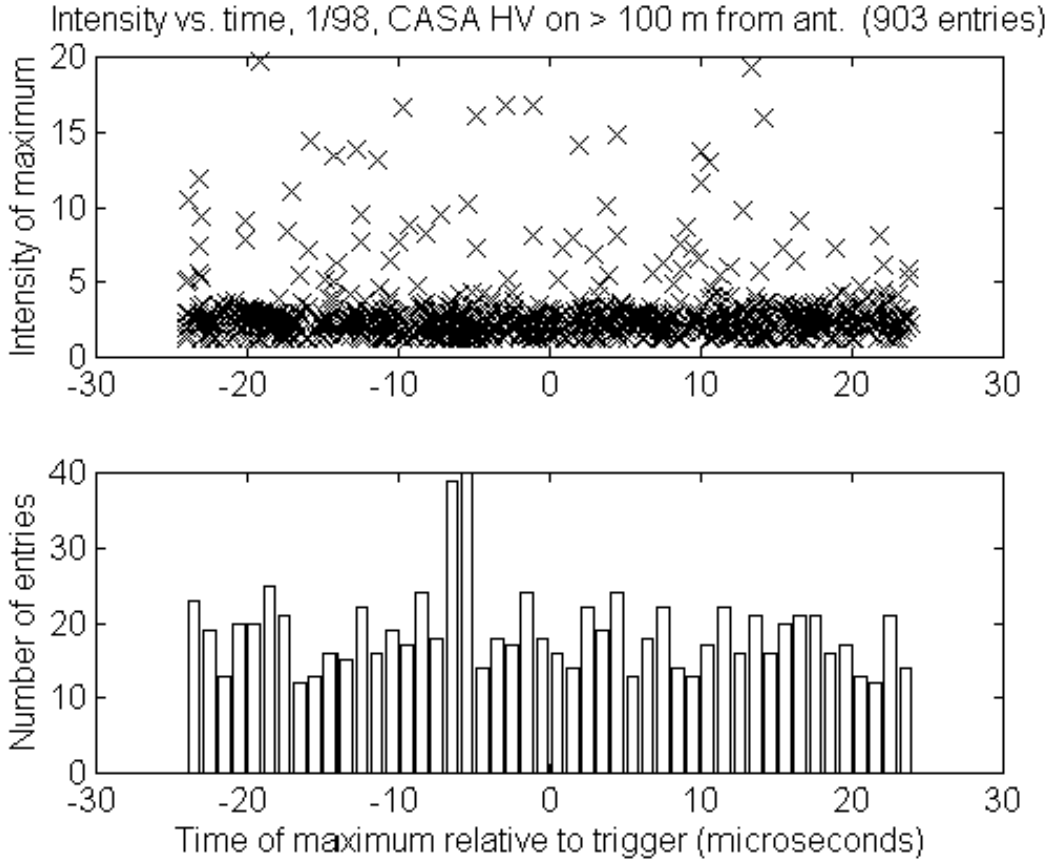}} \caption{Top
panel: intensity-vs.-time plot for maxima of 903 pulses recorded in
620 triggers in January 1998 with CASA HV disabled for boxes within 100 m of
antenna.  Bottom panel:  time distribution of transients.  All events
recorded with East-West antenna polarization.}
\end{figure}

The RF signals from the shower are expected to arrive
no later than, or at most several hundred nanoseconds before, the
transients associated with CASA operation.  They would propagate
directly from the shower to the antenna, whereas transients from
CASA stations are associated with a longer total path
length from the shower via the CASA station to the antenna. There
will also be some small delay at a CASA station in forming the
trigger request pulse.  Thus, we expect a genuine signal also to
show up around 6--7 $\mu$s before the trigger. 

The time coincidence of the CASA RF transients and the shower signals is a 
significant obstacle to detecting genuine pulses. Therefore, data with CASA HV
on or partially disabled were not used for the present analysis.
As we show below in Secs.~5.3.1--5.3.3, no significant peak is visible
around 6--7 $\mu$s before the trigger for data recorded with CASA HV off. 
The upper limit on the rate of events giving rise to such a peak
can be used to set a limit on RF pulses associated with air showers,
as we demonstrate in Sec.~5.3.4.

\subsection{Estimated upper bounds on broad-band signals}

As mentioned in Sec.~5.1, the following discussion is based on 17.25 active 
hours of data accumulation with EW antenna polarization and 21.12 hours with 
NS antenna polarization. The small duration of this subset of data
limits its sensitivity to RF signals from the shower.

\subsubsection{Criteria used to distinguish noise and signal transients}

The main difficulty associated with pulse detection is that
signal pulses are not easily distinguishable from large spurious
pulses originating from atmospheric discharges. Both air shower
pulses and these background noise pulses can considerably exceed
the average noise level. Several criteria can be used to
distinguish signal pulses from noise. The conventional criteria
of the previous studies have been that the pulse should be (1)
larger than the average noise level by some small specified
amount, (2) time coincident with shower particles, and (3)
bandwidth limited \cite{Allan}. All these criteria were adopted
in this study and one more has been added: The pulse should have
approximately uniform distribution over frequency within its
limited bandwidth [see~(c) below].

We now note the particular criteria used to distinguish signal pulses in this
study.

\begin{figure}[t]
\label{2d plot}
\centerline{\epsfysize = 5 in \epsffile{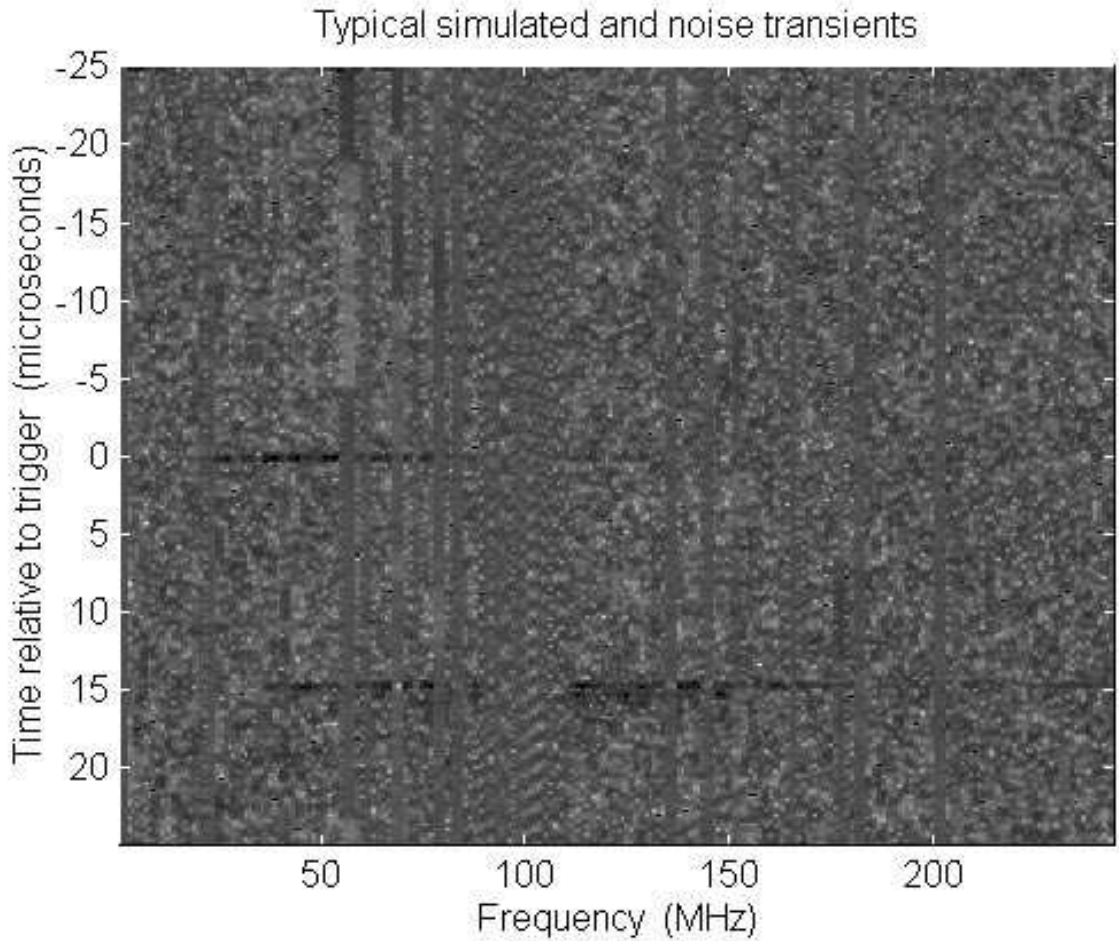}}
\caption{Simulated pulse (horizontal band at $0~\mu s$) and noise
transient (horizontal band at about $15~\mu s$) on a 2-d plot of
intensity (grayscale) vs.\ time (vertical axis) and frequency
(horizontal axis).  Unlike the simulated pulse, the noise
transient contains high frequency components. Grayscale: black
color denotes the highest intensities, white the lowest.
Vertical bands indicate continuous RF sources in the 54--200~MHz range.}
\end{figure}

(a) {\it Pulse magnitude}

Continuous RF interference in each frequency channel was removed by advancing
a moving 1024-ns window in 100~ns steps through the data record to produce a
time-vs.-frequency plot [Sec.~\ref{ss:tfa}] and then using the averaging 
procedure described in Sec.~5.2.
Defining the average signal as 1 (in arbitrary units),
the pulse threshold in each event was taken to be the larger of either
(a) the mean plus three standard deviations, or (b) 1.8 (in the same
units). The former permitted removal of an average noise level; the
latter discriminated against small noise transients.  The final result
was not affected by the choice of the factor 1.8 since the subsequent
analysis [Sec.~\ref{thresh}] used a considerably higher threshold.

(b) {\it Limited bandwidth of pulses}

Broad-band filtering limits the frequency range to 23--250~MHz
[Sec.~\ref{ss:filt}].  The investigation of the simulated
pulses on a 2d-plot of intensity vs.\ time and frequency suggested that, 
unlike some strong noise transients, the signal pulse intensity declines
drastically in the range above approximately 100~MHz (Fig.~11).  This feature
is consistent with theoretical predictions for the shower pulse
spectrum~\cite{SRG}.  It can be chosen as a criterion for
separating noise and signal transients. The ratio of mean intensities
averaged over 23--100~MHz relative to that averaged over 100--250~MHz 
was found to be greater than 1 for all simulated pulses and
smaller than 1 for some noise pulses.

One can consider intensities averaged over the part of the whole 
frequency range. To facilitate separation of signal transients
it is preferable to choose a region where the signal-to-noise
ratio is particularly large. Unfortunately, the whole region from
23~MHz to 100~MHz cannot be effectively used for this purpose.
The 54--82~MHz range is occupied by TV channels 2, 4 and 5, 
which leads to high noise levels. The same is true for the whole FM band
(88--108~MHz) [Sec.~\ref{ss:TVFM}].  However, this is not the case
in the 24--54~MHz range. The noise level in this range is mostly
uniform, and simulated pulse intensities are particularly large
there in comparison with the noise level. The assumption that the
24--54~MHz range provides the best signal-to-noise intensity
ratio has been tested. The results for this range have been
compared with the ones obtained in the 10--54~MHz and 24--86~MHz
regions and were found to be superior. Subsequently, the range of
24--54~MHz was chosen as the main region of investigation.

After that, it was natural to choose the ratio of mean intensities averaged 
over 24--54~MHz relative to that averaged over 55--250~MHz at the moment of
each pulse as a criterion for discriminating noise and signal pulses. For all
simulated pulses this ratio was greater than 1.  All pulses for
which this ratio was smaller than 1 were assumed to be noise
transients and discarded. Also, a ratio parameter threshold other
than 1 can be chosen.  The parameters that provided best results
were found to lie between 1.4 and 1.8 [Section~\ref{res}]. 

(c) {\it Approximately uniform distribution of pulse intensity
over frequency in the 24--54~MHz range}

The most intense noise transients that met criteria (a) and
(b) were found to display a peculiar feature:
Their intensities were concentrated in a small region of
approximately 10~MHz width somewhere in the 24--54~MHz range.
This non-uniformity allowed such pulses to be ruled out. If the
pulse intensity, integrated over {\it any} 9~MHz width window
(in the 24--54~MHz range), was greater than the intensity
integrated over the remaining 21~MHz, then such a non-uniform
pulse was discarded as a noise transient. The reason for choosing
a 9~MHz width window was that most simulated pulses 
passed this test, while many noise transients 
did not.

The effectiveness of these three criteria is illustrated in Table~\ref{pass}.
For simulated pulses that are stronger than 3.43~$\mu$V/m/MHz the combination
of criteria (b), (c) and a very high intensity
threshold provides a $84.1/4.75\approx18$ times increase for their relative 
fraction with respect to noise transients.

\begin{table}[t]
\caption{Fraction of noise pulses passing criterion (a) that can also
pass criteria (c) and (b) with ratio parameter threshold of 1.4, and whose 
maximum intensities are larger than some very high intensity threshold.
Fraction of simulated pulses of different strengths (in $\mu$V/m/MHz), that can
pass criteria (c) and (b) with ratio parameter threshold of 1.4, and whose 
maximum intensities (after they are superimposed on noise) are larger than 
the same high intensity threshold.
The strengths of the pulses are chosen to be stronger than $1/n$ times an
original test pulse amplitude of 13.7~$\mu$V/m/MHz, $n=2,3,4,\ldots,7$.} 
\label{pass}
\begin{center}
\begin{tabular}{|c||c|c|c|c|c|c|c|}
\hline 
{} & {} & \multicolumn{6}{|c|}{Simulated pulses that are stronger than}
 \\ \cline{3-8}
{} & Noise pulses &6.85&4.57&3.43&2.74&2.28&1.96\\ \cline{3-8}
{} & {} & \multicolumn{6}{|c|}{$\mu$V/m/MHz}\\
\hline \hline
North-South & 4.47\% & 94.1\% & 91.7\% & 85.2\% & 66.9\% & 47.7\% & 36.7\%\\
\hline
East-West & 4.75\% & 98.8\% & 93.4\% & 84.1\% & 67.7\% & 48.7\% & 41.4\%\\
\hline
\end{tabular}
\end{center}
\end{table}

\subsubsection{Method outline}

The time coincidence of the pulse with air shower  particles
implies that a signal pulse should be looked for in the range
between $-7$ and $-6~\mu s$ [Sec.~\ref{s:Char}], which will be
referred to in what follows as ``the interesting time bin''. It is important
to note that the aforementioned test criteria are not efficient
if applied only to the pulses found in this time bin. Indeed,
some noise pulses in the bin meet all criteria. Hence, accepting
all the pulses that pass this test would not guarantee that
signal pulses are present among them at all. We developed a
different approach.

The time distribution of noise pulses is assumed  to be uniform.
Hence, the following technique can be adopted. First, the time
distribution histogram of the pulse maxima (from the whole data
run) can be plotted. Then, in the absence of signal pulses, the
number of pulses in a time bin should obey a Poisson distribution
with the average being equal to the average pulse number per bin.
Suppose that an accumulation of pulses in the interesting
time bin $(-7,-6)~\mu s$ is large enough that its probability
according to the Poisson distribution is less than 5\%. Then this
signifies the presence of signal pulses with a confidence level
of 95\%.

Thus, a sufficiently large relative accumulation  of entries in
the interesting time bin was adopted as a key criterion in search
of signal pulses.

\subsubsection{Detailed description}
\label{thresh}

\begin{figure}
\label{fig:17}
\centerline{\epsfysize = 5 in \epsffile{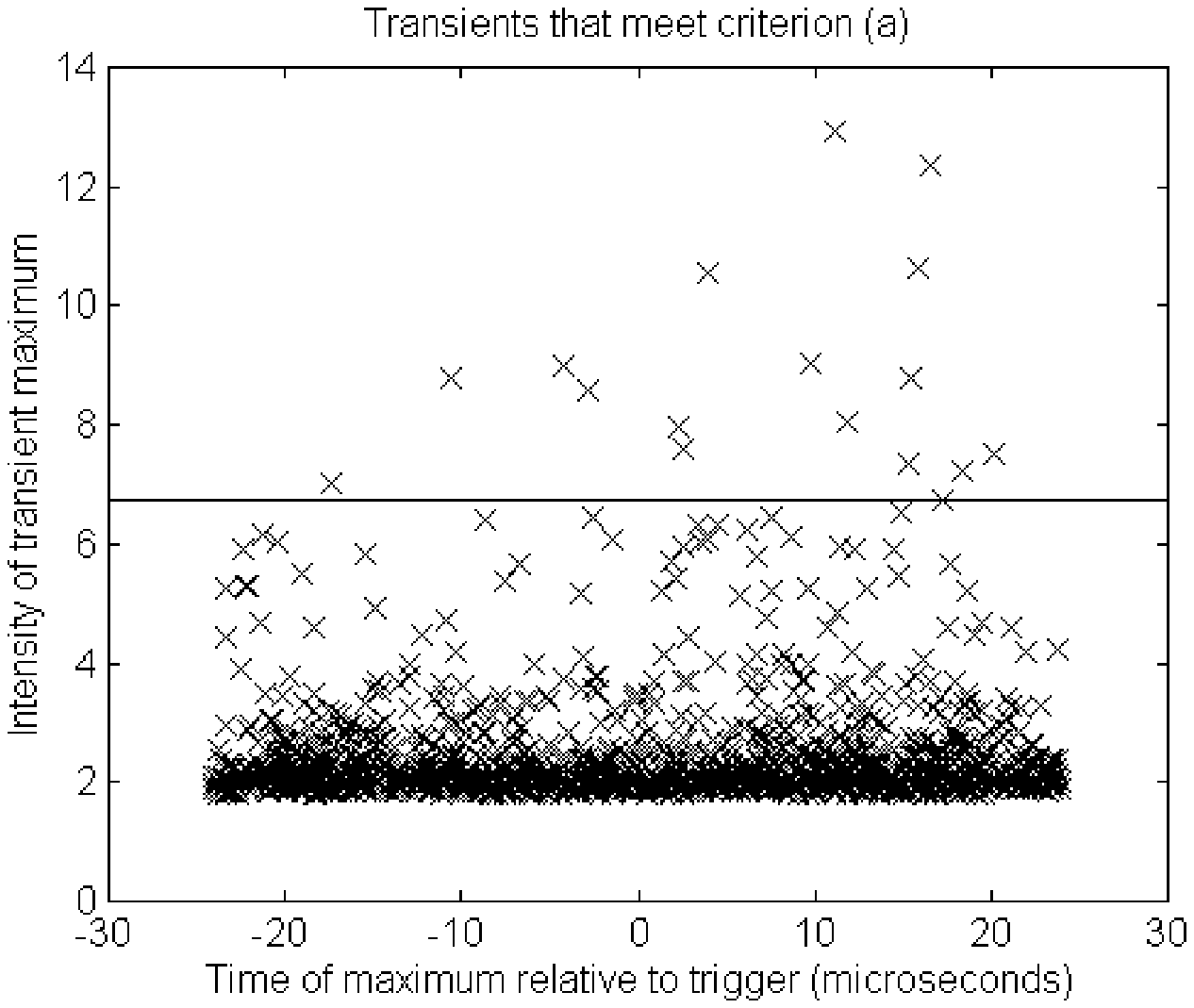}}
\caption{Intensity vs.\ time plot for maxima of transients in 756
files of data with East-West antenna polarization and CASA HV off.
The horizontal line denotes the
threshold taken to lie at the level of the 17th strongest pulse.
Only criterion (a) of Sec.~5.3.1 was employed, resulting in a large (1367)
number of entries.}
\end{figure}

A routine was designed to scan all events in the whole data run
and select the  transients that passed criterion (a) in Sec.~5.3.1.
The ratio parameter (see above) and uniformity parameter [1, if
criterion (c) was satisfied, and 0, if not] were recorded along with time
relative to the trigger and intensity of a transient. The intensity
 vs.\ time plot of pulse maxima selected in this manner for the data
taken with East-West antenna polarization is shown in Fig.~12.

In order to be detectable, signal pulses should be stronger than
the average noise transient, whose intensity equals 2.5 as a result
of the specific criteria imposed.  Hence,
it is reasonable to set the intensity threshold sufficiently high
to enhance the relative accumulation in the $(-7,-6)~\mu s$ time
bin. One can set a high threshold and determine the number of
pulses that pass through it. Divided by number of bins, this
number gives the average pulse number per bin. This, in turn, can
be used as an average value of the Poisson distribution to
determine whether the probability of the observed accumulation in
the interesting bin is less than 5\%.

The effectiveness of this technique depends on the choice of
intensity threshold.  If the threshold is too low, the histogram
includes many noise pulses. Then, the signal pulse accumulation
in the interesting time bin becomes relatively small to be
distinguished from statistical fluctuations. If the threshold is
too high, it may cut out some signal pulses and the remaining
ones may not make up a significant accumulation. The choice of
the optimum threshold is discussed below.

\subparagraph{Optimum intensity threshold.}

Suppose we look for an accumulation of 2 or more entries. Such
an accumulation  is significant (i.e.\ its probability is
$\le5\%$) if the average number of pulses per bin is about 0.355.
Indeed, according to Poisson distribution, $P(\mbox{0
entries})=\exp(-0.355)\approx0.701$, $P(\mbox{1
entry})=0.355\cdot\exp(-0.355) \approx0.249$, so $P(\mbox{$\ge2$
entries})=1-0.701-0.249=0.05$. Since the standard histogram used in
the study contained 48~bins (of 1~$\mu s$ each), the average
value of 0.355 pulses per bin would correspond to a total number
of $0.355\cdot48\approx17$ pulses. These considerations suggest
that the intensity threshold would not have some fixed value but
would correspond to the intensity of the 17th strongest pulse
[see Fig.~12]. Then, the total number of entries in the histogram
is 17 and the average number of entries per bin is about~0.355.
So, 2 or more entries in the interesting bin, if observed, would
have the probability of~5\% and would reveal the presence of
signal pulses with 95\%~CL.

\begin{table}[t]
\caption{Significant accumulation of entries in the interesting
time bin and the corresponding optimum threshold (calculated
according to Poisson distribution).} \label{accum}
\begin{center}
\begin{tabular}{|c|c|c|}
\hline 
\multicolumn{1}{|p{4.5cm}|}{Number of entries in the interesting time 
bin one is looking for}
& \multicolumn{1}{|p{4.5cm}|}
{Average number of entries per
bin at which the probability of this accumulation becomes
$\le5\%$} & \multicolumn{1}{|p{4.5cm}|}{Corresponding
total number of entries in the 48 bin histogram (optimum threshold) }  \\ 
\hline \hline
2&0.355&17\\ 
3&0.817&39\\
4&1.366&65\\
5&1.970&94\\
6&2.613&125\\
7&3.285&157\\
8&3.980&191\\
9&4.695&225\\
10&5.425&260\\
11&6.170&296\\
12&6.924&332\\
13&7.690&369\\
14&8.464&406\\
15&9.245&443\\
16&10.035&481\\
17&10.832&519\\
18&11.635&558\\
19&12.443&597\\
20&13.256&636\\
\hline
\end{tabular}
\end{center}
\end{table}

Naturally, the probability of 3 or more entries would be even
smaller and approximately equal to $P(\mbox{$\ge3$
entries})=P(\mbox{$\ge2$ entries})-P(\mbox{2
entries})=0.05-\frac{0.355^2}{2!}\exp(-0.355) \\ \approx0.006$. That is,
if such an accumulation were observed, the presence of signal
pulses could be claimed with a confidence level of 99.4\%. However,
in reality it is unlikely that signal pulses could be detected
with such a confidence level. To improve the detection chances,
it is advantageous to lower the threshold to the 39th strongest
pulse. This sets the average number of entries per bin to
$39/48\approx0.813$ and makes the probability of 3 entries equal
to~5\%. The threshold of the 39th strongest pulse is the lowest
one that still guarantees a confidence level of at least 95\%.

Thus, the convenient threshold depends on the significant
accumulation number one is looking for. For any particular
accumulation, one can choose the optimum threshold which would
guarantee a confidence level of at least 95\%. The significant
accumulation number and the corresponding optimum threshold can
be found in Table~\ref{accum}.

\subparagraph{}

This method did not reveal any significant accumulation in the
interesting time bin. So far, however, only criterion (a) and
intensity thresholds were applied to detected pulses. As was
already mentioned in the beginning of this Section, the ratio and
uniformity parameters were recorded for each simulated or noise
pulse. This made it easy to impose criterion (b) with different
ratio parameter thresholds and criterion (c). Their application
should have increased the ratio of signal pulse number to noise
pulse number. Unfortunately, no significant accumulation has been
detected (see Fig.~13). This means that signal pulses are quite
rare, so that the resulting accumulation is not significant.
However, upper limits can be placed on the rate (in h$^{-1}$) of
signal pulses that are stronger than some fixed value. This will
be our aim for the next subsection.

\begin{figure}[t]
\centerline{\includegraphics[width=0.9\textwidth]{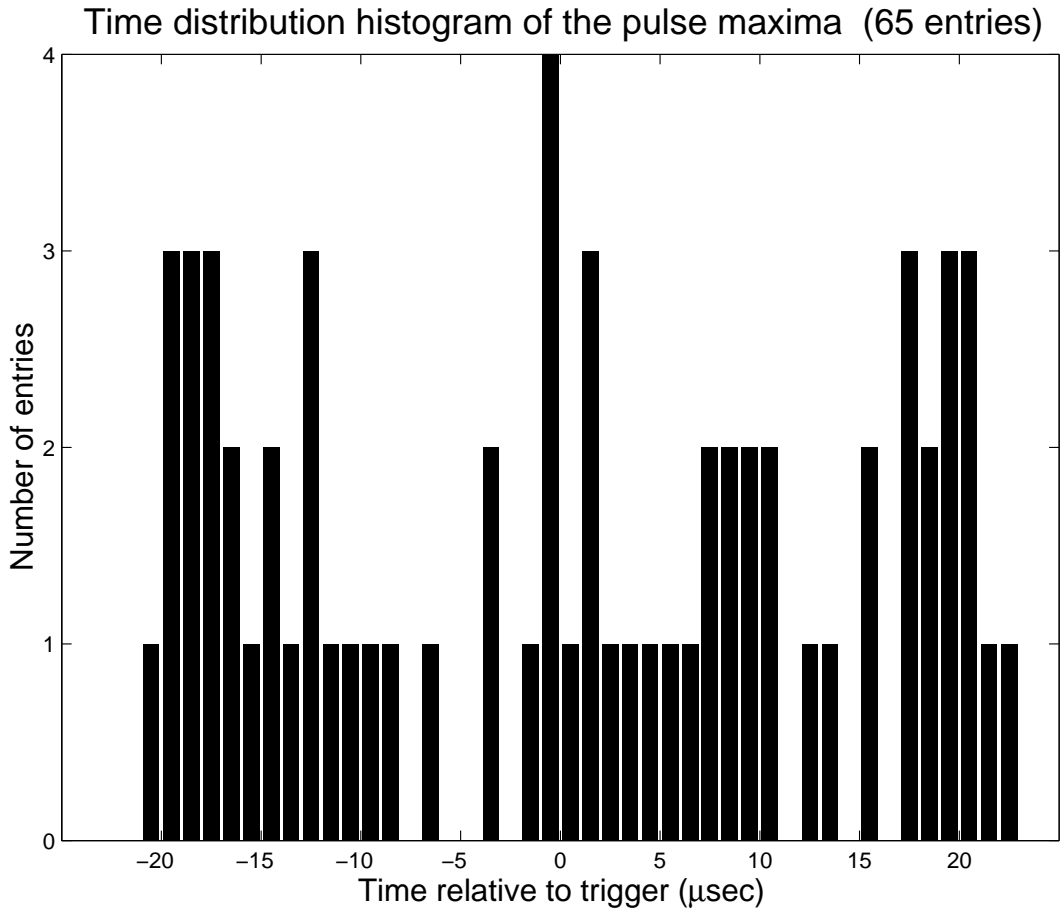}} \caption{
Time distribution histogram for maxima of the 65 strongest pulses  
extracted from 756 files of the EW data with (a), (b) and (c) imposed.
The ratio parameter threshold for criterion (b) is 1.4.
An accumulation of 4 or more pulses in the (--7,--6) $\mu$s
bin would signify the presence of signal pulses. Such an
accumulation was not detected.}
\end{figure}

\subsubsection{Upper limits on the rate}
\label{res}

The total number $n$ of observed transients in the interesting time bin
consists of both noise and signal events. The former are Poisson-distributed 
with known average $\mu$. The latter obey a binomial distribution with a 
known probability $P$ to pass thresholds and an unknown total number $N$ of 
signal events contained in the run. Using a unified approach to the 
statistical analysis of small signals~\cite{FC}, we construct 95\% confidence
belts for unknown $N$. For any choice of imposed thresholds, the lower end of
confidence intervals is 0, i.e.\ only upper limit on the total number of 
signals in the run can be placed.

\subparagraph{Probability $P_s(k)$ that the interesting bin contains $k$ 
signal events.}

Formula (1) [Section 3] shows that ${\cal E}_\nu$ is directly proportional to
primary energy $E_p$.  The differential rate of primaries is known to fall
approximately as $1/E_p^3$ \cite{REF}.  Similarly, the differential rate of
${\cal E}_\nu$'s should be proportional to $1/{\cal E}_\nu^3$.

Simulated pulses of different strength ${\cal E}_\nu$ greater
than some ${\cal E}_\nu^0$ can be added to each event of the data
run.  To simulate the expected rate for ${\cal
E}_\nu$, a Monte-Carlo simulation was performed in such a way that
the number of added simulated pulses with strength ${\cal E}_\nu$
falls as $1/{\cal E}_\nu^3$. 

Each event underwent the averaging procedure described in Sec.~5.2. 
The averaged intensities of simulated pulses were compared to the values of
the 17th (or 39th, 65th, etc.) strongest noise pulse of the
initial data (without added simulated pulses). The fraction of
simulated signal pulses that were higher than this threshold gives the
probability $P$ for a signal pulse with strength greater than
some ${\cal E}_\nu^0$ to exceed this threshold. We determined the value of 
$P$ as a function of ${\cal E}_\nu^0$, the intensity threshold, and the set 
of 
criteria ((a), (b),(c), or some combination of them) imposed on both noise 
and signal pulses. Then, these values were used to make an analytical 
estimate of the probability to detect a significant accumulation
number of entries in the interesting bin.

By making an assumption for the total number of signal pulses $N$
during the run and using the binomial distribution, one can
calculate the probability that $k$ out of these $N$ signal pulses exceed
the threshold: $P_s(k)=C(N,k)\,P^k\,(1-P)^{N-k}$, where
$C(N,k)$ is the standard binomial coefficient. This
probability not only depends on ${\cal E}_\nu^0$, the intensity threshold, and 
the set of imposed criteria but also on the aforementioned assumption for
the number $N$. 

\subparagraph{Determining upper limits $N^{up}$.}

Suppose, for instance, that we are looking for 3 or more
significant entries in a bin. 

First, we impose a particular set of criteria on all pulses in the run,
pick the 39 strongest of them and determine the number $n$ of entries in the 
interesting bin. 39 pulses in the histogram correspond to an average
$\mu=39/48\approx0.813$ entries per bin. The Poisson
distribution $P_n(k)=\mu^k e^{-\mu}/k!$ determines the probability to have 
$k$ pulses in a bin.

Second, we set the intensity threshold at the level
of the 39th pulse. Then we determine $P_s(k)$ for the same set of imposed 
criteria, this intensity threshold and any total number $N$ of signal 
transients in the run.

Consider the construction of an acceptance interval of $n$ values for some 
fixed total number $N$ of signal transients in the run. $n$ entries in the 
interesting bin can be the result of different combinations of noise and 
signal entries. We calculate the 
probability to detect $n$ entries in the bin as 
$$
P(n|N)=\sum_{k=0}^n{P_s(k)P_n(n-k)}=e^{-\mu} 
\sum_{k=0}^n{C(N,k)\,P^k\,(1-P)^{N-k}}\mu^{n-k}/(n-k)!
$$
Take some values of $n$ and $N$, for example, $n=n_0$ and $N=N_0$. The 
probability $P(n_0|N_0)$ might be small but not so small with respect to 
$P(n_0|N_{best})$, where $N_{best}$ is such an alternate hypothesis which 
maximizes $P(n_0|N)$. The ratio $R=P(n_0|N_0)/P(n_0|N_{best})$ is the basis
of the ordering principle outlined in~\cite{FC}.

For any $N$ we add values of $n$ into the acceptance interval in decreasing 
order of $R$. As soon as the sum of $P(n|N)$ exceeds the confidence level of 
0.95, the acceptance interval is completed. The confidence belts that can be 
constructed using this procedure are shown in Fig.~14. For any observed 
number $n$ of entries in the interesting bin, these belts provide 
intervals for allowed values of $N$. For any choice of imposed criteria and 
for all values of $n$ measured in the experiment, the lower end of the 
intervals was 0. Thus, the actual signal transients were not detected and only the upper limit $N^{up}$ on their number during the run 
can be placed.

\begin{figure}[t]
\centerline{\includegraphics[width=0.9\textwidth]{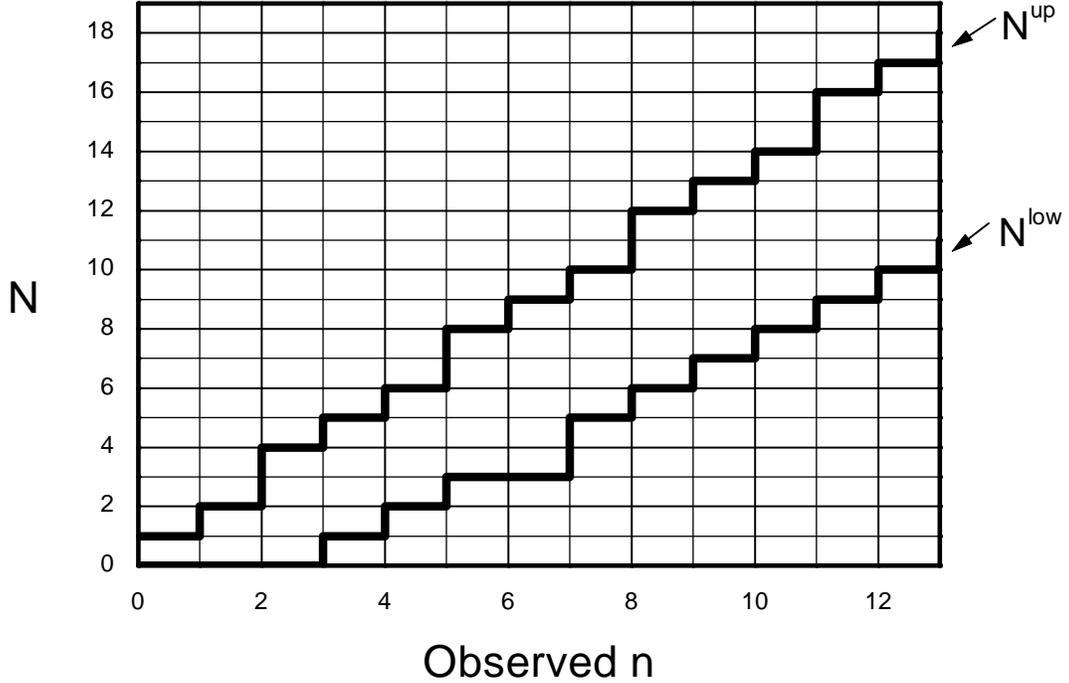}}
\caption{Confidence belt based on ordering principle of reference~\cite{FC}, 
for 95\% CL intervals for unknown total number $N$ of signal pulses in the 
run. The probability $P=0.841$ for a signal pulse to go through a set of  
cuts, and the Poisson background mean $\mu=65/48=1.354$, are the parameters 
of the plot. The presence of $n\ge4$ entries in the interesting time bin 
from the total of 65 entries in the histogram would signify the presence 
of signal pulses in the run [see Table~10]. 
This is also reflected in this Figure as for $n\ge4$ $N^{low}$ becomes 
nonzero.
$P=0.841$ corresponds
to the probability for signal pulses stronger than 1/4 the original test pulse
amplitude to pass criteria (b), (c),
and an intensity threshold at the level of the 65th strongest pulse of 
the EW data [see Table~9]. The ratio parameter threshold for criterion (b) is 1.4.
Only one of the 65 strongest pulses was observed in the interesting time bin
[see Fig.~13],
implying the upper limit $N^{up}=2$ 
(see also the corresponding dot in 
Fig.~15(b) at ${\cal E}_\nu^0=3.43~\mu$V/m/MHz).}
\end{figure}

\subparagraph{Upper limit $R^{up}({\cal E}_\nu>{\cal E}_\nu^0)$.}

Divided by the total run time (17.245~hours 
with EW antenna polarization and 21.116~hours with NS polarization), $N^{up}$
gives the upper limit on the rate of signal pulses of 
strength greater than ${\cal E}_\nu^0$. We will denote it
$R^{up}({\cal E}_\nu>{\cal E}_\nu^0)$. Of course, one can obtain
different values of $R^{up}({\cal E}_\nu>{\cal E}_\nu^0)$ when looking for
a different significant number of entries in the interesting bin. We searched
for the accumulation values from 2 to 20 in the bin $(-7,-6)~\mu s$.
The lowest $R^{up}({\cal E}_\nu>{\cal E}_\nu^0)$ gives the most
stringent upper limit on the rate. Its value  also depends on criteria 
imposed on both signal and noise pulses. The best results were achieved when
all three criteria (a), (b) and (c) were employed. 
The ratio threshold parameter for criterion (b) was tested in the region from
1 to 2.4 with step 0.2. The thresholds that give the best results were found 
to be 1.4 for EW and both 1.6 and 1.8 for NS data. Compared to the analysis 
with neither (b) nor (c) employed,  these threshold parameters provide up to
28\% lower upper limits. The resulting values of $R^{up}({\cal E}_\nu>{\cal 
E}_\nu^0)$ are shown as dots in Fig.~15 for several values of ${\cal 
E}_\nu^0$ for the data taken with both East-West and North-South 
polarizations. Signal pulses stronger than 4.57~$\mu$V/m/MHz are expected to 
be quite rare and were definitely absent from the NS data, just like pulses 
stronger than 6.85~$\mu$V/m/MHz were absent from the EW data. Zero upper 
limits on the rates of these events are indicators of these facts.

\begin{figure}
\centerline{\includegraphics[width=0.9\textwidth]{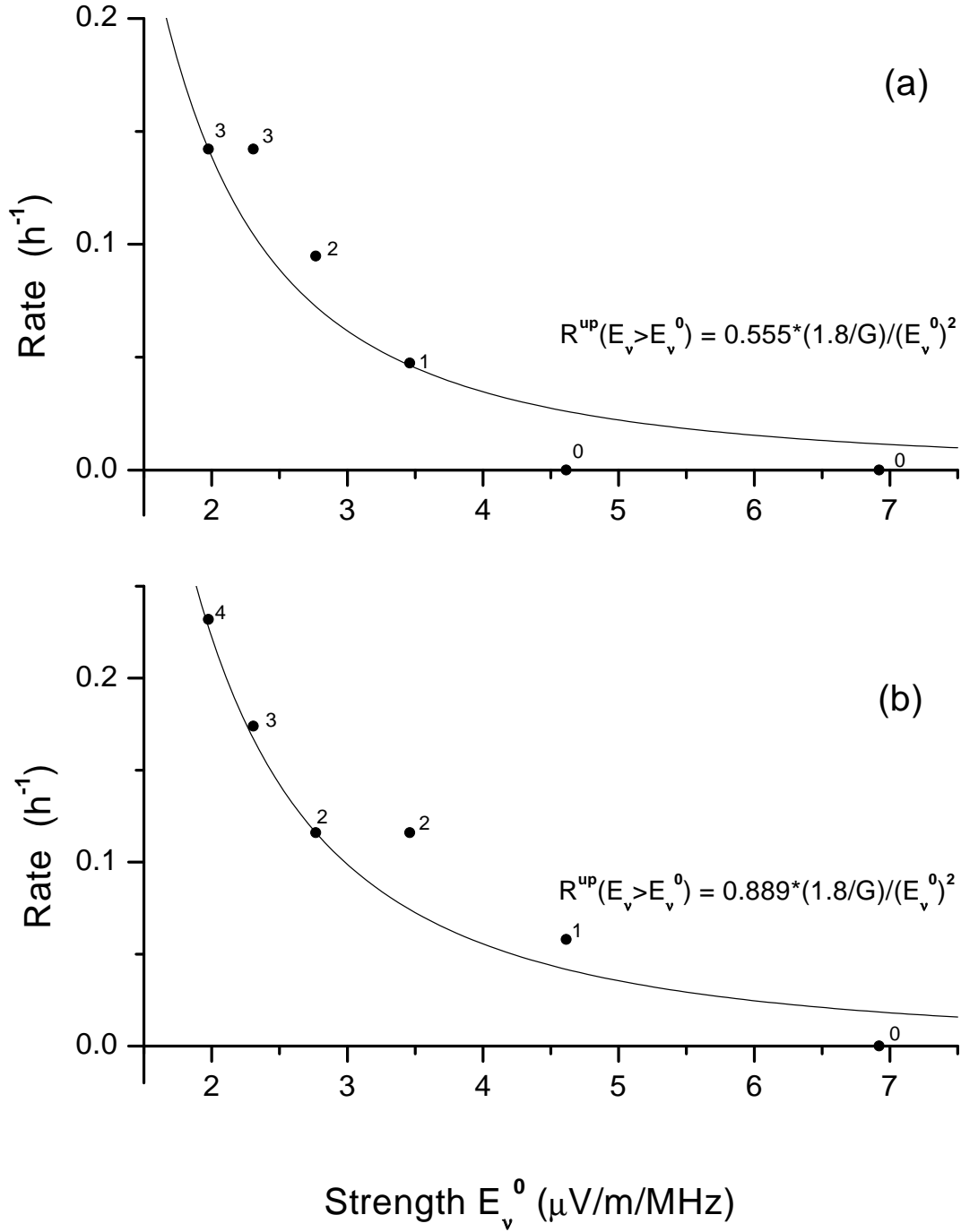}}
\caption{Upper limit $R^{up}({\cal E}_\nu>{\cal E}_\nu^0)$ on the
rate of signal pulses stronger than a fixed strength ${\cal E}_\nu^0$.  
Plotted dots show the upper limit for antenna
system gain $G=1.8$. Small numbers near the dots indicate $N^{up}$,
the upper limit on the number of the signal events in the run.
The dots were plotted for the pulses that are stronger than $1/n$ times the 
original test pulse amplitude of 13.7~$\mu$V/m/MHz, $n=2,3,\ldots,7$. 
The solid line represents the lowest upper limit corresponding to them.  Data
were taken with (a) North-South (21.12~hours) and (b) East-West (17.25~hours)
antenna polarization.}  
\end{figure}

As was mentioned above, the differential rate $R({\cal E}_\nu)$ should be
proportional to $1/{\cal E}_\nu^3$. Then, the rate of signal
pulses of the strength greater than ${\cal E}_\nu^0$ is $$
R({\cal E}_\nu>{\cal E}_\nu^0)=\int\limits_{{\cal
E}_\nu^0}^{+\infty} R({\cal E}_\nu)\,d{\cal
E}_\nu\propto\int\limits_{{\cal E}_\nu^0}^{+\infty}
\frac{1}{{\cal E}_\nu^3}\,d{\cal E}_\nu\propto
\frac{1}{({\cal E}_\nu^0)^2} $$ i.e., proportional to $1/({\cal E}_\nu^0)^2$.
Hence, the lowest curve $c/({\cal E}_\nu^0)^2$ passing through one of the
nonzero dots (solid line in Fig.~15) represents the strictest upper limit on
this rate.  The value of the coefficient $c$ was found to be 0.555 for the NS
data and 0.889 for the EW data:
\beq
R^{up}({\cal E}_{\nu NS}>{\cal E}_\nu^0)=0.555/
({\cal E}_\nu^0)^2 \ \ {\rm h}^{-1},
\label{eqn:uplimNS} 
\eeq 
\beq
R^{up}({\cal E}_{\nu EW}>{\cal E}_\nu^0)=0.889/
({\cal E}_\nu^0)^2 \ \ {\rm h}^{-1},
\label{eqn:uplimEW} 
\eeq  
where ${\cal E}_\nu^0$ is in $\mu$V/m/MHz.  These results should be interpreted
as the upper limits on the rates of events where the North-South (or East-West)
projection of the signal pulse was greater than some value. We
will use them in Appendix A for sensitivity calculations.

\subsection{Discussion and summary}

The small duration of the present subset of data prevents us from confirming
or refuting previous claims for shower pulses.  As mentioned, this subset was
taken with CASA HV off, and was initially collected in order to check
indications of a signal which occurred with CASA HV on.  Using the estimate of
2.25 strong signal pulses per day [Sec.~4.4.4], one expects about three or four
detectable shower pulses during 38.36~hours of observation.
Three or four detectable particles that might give a strong signal during
the experiment would be too scanty an amount to be detected over the noise
background, which closely mimics genuine signals. 

\begin{figure}
\centerline{\epsfysize = 6.55 in \epsffile{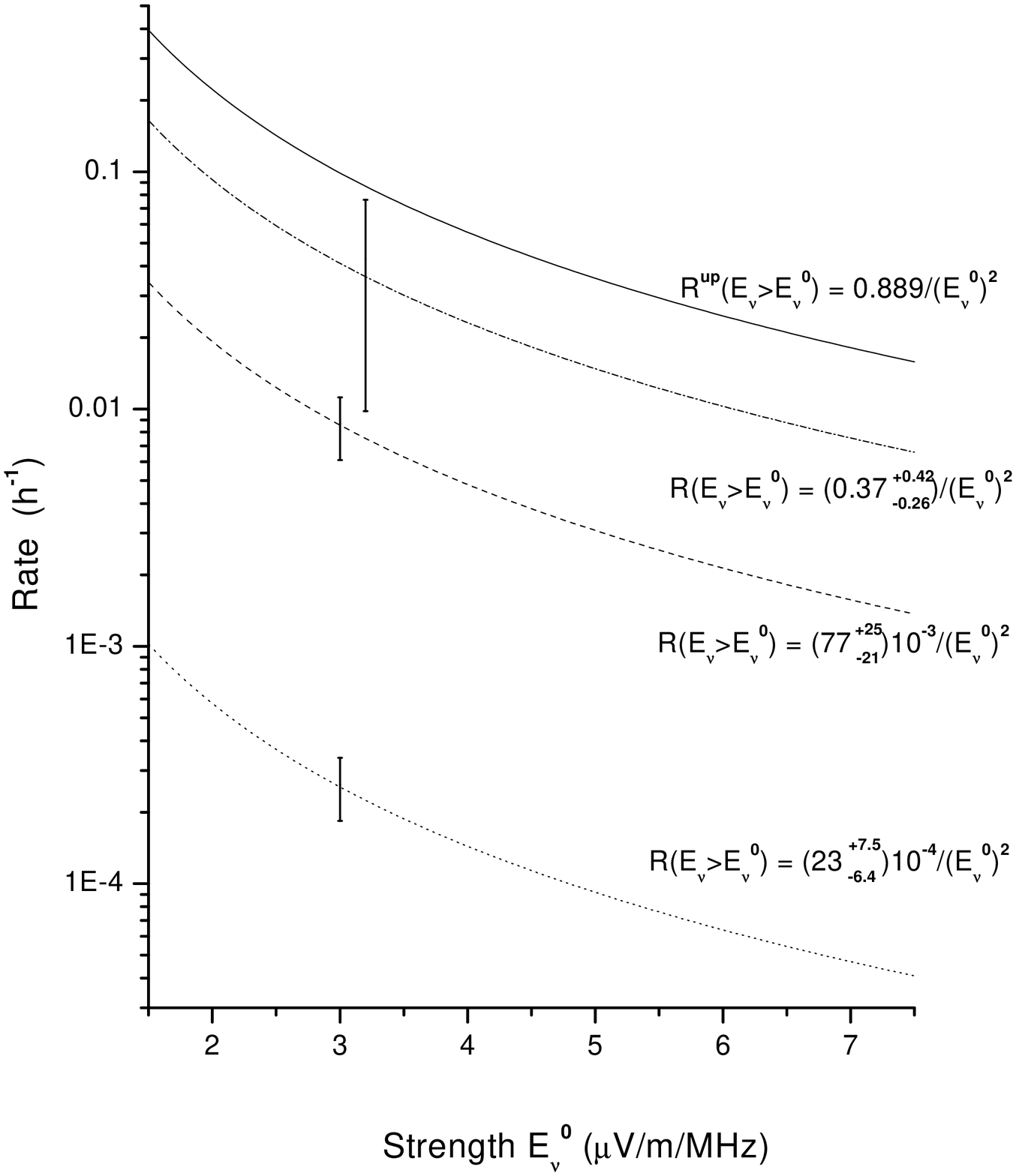}} 
\caption{Rate of events at the CASA detection area in which the East-West
projection of the signal pulse is greater than some ${\cal E}_\nu^0$.  The
lower three curves show the {\it predicted} rates based on different values of
$s$:  $s = 20 \pm 9.5$ (dash-dotted line, the original Haverah Park result
\cite{Allan} based on $\sim$100 detected shower pulses); $s = 9.2 \pm 1.4$
(dashed line, the Soviet group result \cite{Atrash} based on $\sim$1000 
pulses), and $s = 1.6 \pm 0.24$ (dotted line, the updated Haverah Park result
\cite{Atrash} based on $\sim$1000 pulses).  The uncertainty in the latter two
values is 15\%~\cite{Atrash}. We evaluated the uncertainty in the former as
being $\sqrt{10} \cdot 15\% \approx 47\%$.  In this logarithmic plot the error
bars have constant lengths along their corresponding curves.
The top curve is the upper limit in this paper which corresponds 
to $s=31$.  These four curves are determined under the assumption 
of a transverse current radiation mechanism~\cite{KL}.}
\end{figure}

\begin{figure}
\centerline{\epsfysize = 6.55 in \epsffile{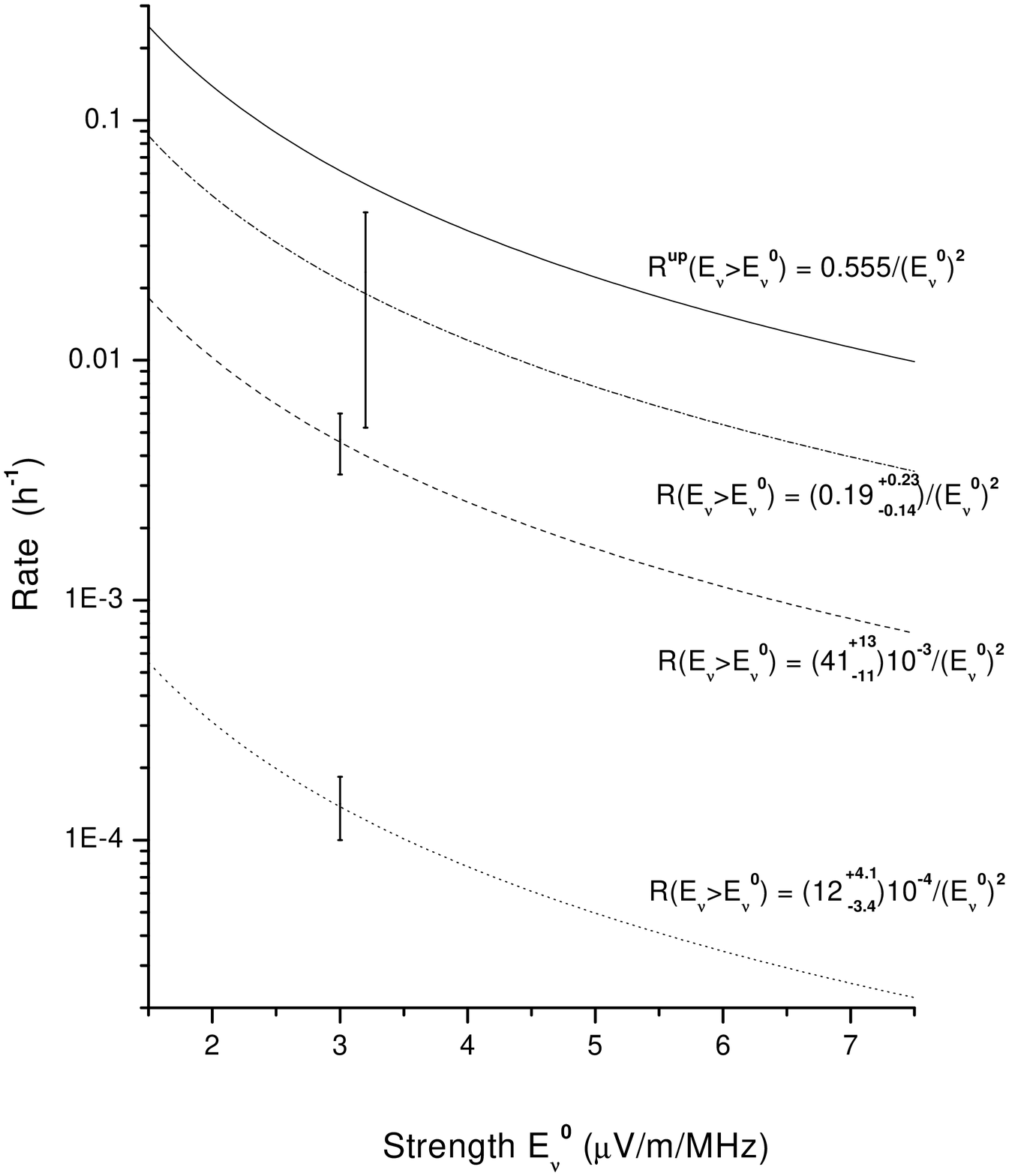}} 
\caption{Rate of events at the CASA detection area in which the North-South
projection of the signal pulse is greater than some ${\cal E}_\nu^0$.  The
lower three curves show the {\it predicted} rates based on different values of
$s$:  $s=20\pm9.5$ (dash-dotted line, the original Haverah Park result
\cite{Allan} based on $\sim$100 detected shower pulses); $s=9.2\pm1.4$
(dashed line, the Soviet group result~\cite{Atrash} based on $\sim$1000 
pulses), and $s=1.6\pm0.24$ (dotted line, the updated Haverah Park 
result~\cite{Atrash} based on $\sim$1000
pulses). The uncertainty in the latter two values is 15\%~\cite{Atrash}. We 
evaluated the uncertainty in the former as being $\sqrt{10}\cdot15\% \approx 
47\%$. In this logarithmic plot the error bars have constant lengths along 
their corresponding curves.  The top curve is the upper limit in this paper
which corresponds to $s=34$.  These four curves are determined under the
assumption of a particular radiation mechanism discussed in Appendix~A.}
\end{figure}

Nonetheless, we were able to place upper limits on the integral rate of 
signal pulses. Under certain simplifying assumptions these upper limits can 
be used for evaluating the numerical calibrating factor $s$. This factor was 
set to 20 in Eq.~(\ref{eqn:E}) but there remained an uncertainty regarding 
its value. In fact, subsequent reports claimed values as small as~1.6. 
We could only set upper limits on $s$: $s<31$ from the EW and $s<34$ from the
NS data [see Appendix~A]. The former was obtained under the assumption 
of the transverse current radiation mechanism, the latter under an 
alternative radiation mechanism discussed in Appendix~A.
Figs.~16 and 17 compare our results with those of previous experiments.

Had additional observation time with CASA boxes disabled been available, more
signal pulses and more noise transients would be recorded. Would it be easier
to prove the presence of shower signals in the data? Taking into account that
the average signal pulse from a high energy air shower is much stronger than 
the average noise pulse, the content of the most intense transients (those 
stronger than 17th strongest, 39th, etc.) would shift in favor of more signal
pulses. Thus, additional running time with CASA HV off would have allowed one
to at least place a stricter upper limit on the rate of shower pulses, if not
to detect them. 

\section{Issues specific to a giant array}

\subsection{Frequency range}

The study of pulse components above 30~MHz remains a useful
restriction in view of the changing RF environment encountered at
lower frequencies.  During years of sunspot minima the whole
range of frequencies above 23~MHz remained relatively clear of
broadcast interference, while as the sunspot numbers increased
toward the end of 1997, daytime interference became particularly
intense from the citizen's band just above 27~MHz.  This
interference typically subsided at sunset.  If observations
below 30~MHz are contemplated, they would be most
useful during years of expected sunspot minima (e.g., 2006-7), to
minimize effects of long-distance ionospheric reflections.

\subsection{Number and spacing of receiving sites}

It is expected \cite{Allan} that signals above 30~MHz decrease
rapidly as a function of distance between the antenna and the
shower core's closest approach.  Thus, we expect that in the
Auger array, with spacing of 1.5 km between sites in a hexagonal
array, RF stations would have to be distributed with roughly the
same or greater density.  One possibility for minimizing RF
interference from Auger stations would be to place the RF station
at the interstices between them.  This would raise the cost per
station, since it would require microwave communication with
Auger sites and auxiliary sources of power.

At each station it may be helpful to have two antennas, one
registering pulses of east-west polarization and one for
north-south polarization.  Differential signals as well as
individual ones should be recorded.  Coincidences among several
stations may be associated with particularly large showers.
Frequency-dependent antenna phase response should be modeled or measured, as
noted in Sec.\ 4.4.8.

\subsection{Digitization requirements}

Previous work by one of us (J.F.W.) involved detection of electromagnetic
pulses, including those possibly produced by cosmic-ray-induced electromagnetic
discharges, with frequencies in the 30 -- 100~MHz range.  Part of this work
included building hardware for self-triggering on short duration wide-band RF
pulses.  Many of the pulse identification, fast-digitization and memory
problems were identical to those for pulse detection at CASA/MIA.
Time-frequency plots were obtained similar to those one would generate in a
survey at CASA/MIA.  Similar requirements also are encountered for digitization
of data from the KamLAND Experiment \cite{KAM}.

Our experience with the present system indicates the need for
expanded dynamic range if continuous-wave sources of RF
interference (such as FM and TV broadcast stations) are to be
eliminated digitally.  Thus, one needs at least 10-bit and
probably 12-bit range, with a digitization rate of at least 
400~MHz so as to be sensitive to frequencies up to 200~MHz. 
Although the expected signal is likely to be concentrated at lower
frequencies (probably below 100~MHz), the expanded frequency
range has proved useful in distinguishing expected signals from
other transients.

\subsection{Stand-alone trigger}

Our results do not indicate that a trigger based on RF signals
alone can yield useful correlations with air shower events.  This
result may be specific to the location of the CASA/MIA array; such
a trigger may be less subject to noise at a remote location such
as the Southern Hemisphere Auger site.  An RF survey performed
there would be useful in determining the utility of such a
trigger.  At such a site, more free from man-made noise than the
CASA/MIA site, one would have to perform further studies allowing
discrimination between random triggers (such as those induced by
atmospheric discharges) and those induced by air showers.  One
would also attempt to detect galactic noise as a further
indication that the site was sufficiently quiet.

\subsection{Integration into the data stream}

The data of CASA/MIA and that of the RF detection experiment were
only integrated off-line.  Any further studies should allow for
simultaneous acquisition of both sets of data.  Since the Auger
project proposes to use microwave communication between stations,
this same link should be considered for communicating RF signal
acquisition results to a central data stream.

\subsection{Status of GHz detection}

David Wilkinson, who visited the University of Chicago during the
spring of 1995, has proposed looking into the power radiated at
frequencies of several GHz, where new opportunities exist
associated with the availability of low-noise receivers.  These
techniques have now been implemented in the RICE project
\cite{RICE}, which seeks to detect pulses with frequency
components around 250~MHz in Antarctic polar ice.

\subsection{Other options}

Dispersion between arrival times of GPS signals on two different
frequencies may serve as a useful monitor of air shower
activity.  The possibility of correlation of large showers with
such dispersion events could be investigated.

It may be possible to monitor commercial broadcast signals in the
54 - 216~MHz range to detect momentary enhancements associated
with large showers, in the same sense that meteor showers produce
such enhancements.  Television channels for which no nearby
stations exist offer one possibility.
The data taken at CASA/MIA have not yet been analyzed in terms of such
enhancements, but represent a potential source of information.
In Table~\ref{TVstat} we note the locations of TV stations
broadcasting on VHF channels {\it other} than those assigned to
the Salt Lake City metropolitan area within 400~km of Dugway.
These are channels 3 (60--66~MHz), 6 (82--88~MHz), 8 (180--186~MHz), 
10 (192-198~MHz), and 12 (204-210~MHz).  The availability
of at least two stations on Channel 3 and three on Channel 6 at
greatly differing headings indicates that this method may have
some promise.

\begin{table}
\caption{Television stations within 400~km of Dugway on channels
not assigned to the Salt Lake City metropolitan area.
Ref.~\cite{TV} also lists four ``new'' (unidentified) stations on
Channel 3 for Price, UT (distance $\simeq$ 160~km, heading
$\simeq$ 110 degrees) and two on Channel 12 for Logan, UT
(distance 178 km, heading 14 degrees).}
\label{TVstat}
\begin{center}
\begin{tabular}{l c r r c c} \hline
Call  &   Location      & Channel & Power & Distance from & Heading  \\
sign  &                 &         & (kW) & Dugway (km)    &
(degrees) \\ \hline
KBJN  &    Ely, NV      &    3    & 100  &      212       &    239   \\
KIDK  & Idaho Falls, ID &    3    & 100  &      363       &     1    \\
KBNY  &    Ely, NV      &    6    & 100  &      212       &    239   \\
KPVI  & Pocatello, ID   &    6    & 100  &      301       &     6    \\
KBCJ  &   Vernal, UT    &    6    & 83.2 &      306       &     87   \\
KIFI-TV & Idaho Falls, ID &  8    & 316  &      363       &     1    \\
KENV  &    Elko, NV     &   10    & 3.09 &      272       &    280   \\
KISU-TV & Pocatello, ID &   10    & 123  &      301       &     6    \\
KUSG  & St.~George, UT  &   12    &  10  &      359       &
191   \\ \hline
\end{tabular}
\end{center}
\end{table}

Radar detection of showers offers another exciting possibility
\cite{PG}. This method resembles the use of
distant fixed VHF stations for generating reflections off
showers, but allows for a more carefully controlled
environment.

\section{Conclusions}

A prototype system for the detection of radio-frequency (RF)
pulses associated with extensive air showers of cosmic rays was
tested at the Chicago Air Shower Array and Michigan Muon Array
(CASA/MIA) in Dugway, Utah.  This system was under
consideration for use in conjunction with the Pierre Auger
Project, which seeks to study showers with energies above
$10^{19}$~eV.

The system utilized a trigger based on the coincidence of 7 out
of 8 buried muon detectors around the periphery of the CASA
array.  Transients were indeed detected in conjunction with large
showers, but they were identified as arising from the CASA
modules themselves, most likely from the electronics generating
trigger request (TRQ) pulses.  Such transients could be
eliminated when the high voltage (HV) on CASA phototubes was
turned off; in such cases the muon trigger continued to
function.  
By comparing upper limits on detected transients with
simulated pulses, it was possible to place upper bounds on
the rate of detection of RF pulses of various intensities.  These
upper bounds are summarized in Fig.~15; they typically involve
rates of one every few hours for the largest field strengths
claimed in the literature \cite{Atrash}.  
Based on our estimates, it is unlikely that the
present experiment can reach the sensitivity limits of the
Haverah Park results, which reported lower field strengths in
their latest work \cite{Atrash}.

A number of lessons have been learned from the present exercise.
These are probably most relevant for any installation
contemplated in conjunction with the proposed Auger project
\cite{Auger}.

(1) One must take special care to survey transients produced by
components of the array.  For the Auger detector, one must install
one or more antennae close to the proposed \v{C}erenkov detectors
and their associated digitizers, and study the response to
artificially induced signals.
Based on our experience, in which the present bounds are based on a
small subset (38 hours) of the total data sample (600 hours), one should
focus as soon as possible on a configuration in which usable data can be
gathered.

(2) The method of communication between Auger modules will affect
what form of RF detection is feasible.  If radio
links employing microwave ($> 800$~MHz) frequencies are used, the
present system will not be as seriously compromised as it
apparently was by the communication system used at CASA/MIA. On
the other hand, detection of RF signals above 1~GHz will suffer
interference from such a system.

(3) Consideration should be given to placement of antennas at
sites sufficiently far from surface detector modules that pulses
from these modules do not constitute a serious source of
interference.  In the Auger case, the modules are arranged in a
triangular array with 1.5~km spacing, so that the maximum spacing
between an antenna inside the array and any module could be as
large as $1.5/\sqrt{3} \simeq 0.87$~km if the antenna is placed
at an equal distance from the three closest modules.

(4) Coincidence between RF signals detected at several antennas
is desirable, as was found in the earliest experiments \cite{Chac}.

(5) The digital filtering algorithms employed in the present
study, although not yet pushed to their optimal efficiencies,
appear to be limited by the 8-bit dynamic range employed in
detection using a digitizing oscilloscope. Consideration should
be given to a system with larger dynamic range, at least 10-bit
but preferably 12-bit.

(6) One can probably afford to economize by reducing the sampling rate,
certainly to 500 MSa/s but perhaps as low as 200 MSa/s, since
transients are expected to have their main frequency components
below 100 MHz, and by reducing the active sampling window from
the present value of 50 $\mu$s to a lower value, depending on the
geometry of the array.

(7) The frequency range studied in the present work (23--200~MHz)
will be more useful if continuous RF sources,
such as FM and television stations, are much weaker than
they were at the Dugway site. This is a possibility in the remote
Argentine site at which the first Auger array is to be
constructed \cite{Auger}; a survey of field strengths there would
be desirable.

(8) One should take data simultaneously with two antennas
polarized in perpendicular (EW and NS) directions.  This allows
the determination of two components of the electric vector, not
just one of its projections.

Although no RF signals have been detected in conjunction with
CASA/MIA events, the present study has revealed a number of
useful criteria for future experiments of similar type.  It is
hoped that a prototype at an Auger site will further focus these
criteria.

\section*{Acknowledgments}

It is a pleasure to thank Mike Cassidy, Jim Cronin, Brian Fick, Lucy Fortson,
Joe Fowler, Rachel Gall, Brian Newport, Rene Ong, Scott Oser,
Daniel F. Sullivan, Fritz Toevs, Kort Travis, and Augustine Urbas
for collaboration and support on various aspects of this experiment.
Thanks are also due to Bruce Allen, Dave Besson, Maurice Givens, Peter 
Gorham,
Kenny Gross,
Dick Gustafson, Gerard Jendraszkiewicz, Larry Jones, Dave Peterson, John
Ralston, Leslie Rosenberg, David Saltzberg,
Dave Smith, M. Teshima, Stephan Wegerich, and David Wilkinson for
useful discussions.  This work was supported in part by the Enrico Fermi
Institute, the Louis Block Fund, and the Physics Department of the University
of Chicago and in part by the U. S. Department of Energy under
Grant No.~DE FG02 90ER40560.

\section*{Appendix A:  Sensitivity Calculation}

One of the aims of this study was to evaluate the antenna
response as a function  of shower parameters. The dependence on
$R$, $\theta$, $\alpha$, $E_p$ has been established in \cite{Allan}
(Eq.~(1)). However, there remained an uncertainty as regards
the calibrating factor $s$, for which experimental data do not
supply an exact value. Indeed, we have quoted at least three
versions of it: first, in Eq.~(1) it is set to 20; second,
the Haverah Park group subsequently reported that ${\cal
E}_\nu^N=0.6$, while the Soviet group claimed the value of 3.4
\cite{Atrash}. To make the connection between ${\cal E}_\nu^N$ and $s$, let
us note that $s \simeq {\cal E}_\nu^N/\exp(-1)$. Thus, we can infer that
the Haverah Park group updated Eq.~(1) to $s=1.6$, and
the Soviet group to $s=9.2$.

Ideally, we would like to calculate the calibrating factor
precisely from  the rate $R({\cal E}_{\nu EW}>{\cal E}_\nu^0)$,
but unfortunately this rate cannot be obtained accurately from
experiment; we could only set an upper bound on it (Section
5.3.4). Hence, in this Appendix we will be able to set only an
upper limit on $s$ to get at least some evaluation of the
possible range of its values. We will do it in the following way.

Equation (1) and the exact knowledge of the coefficient $k$ in the
formula for the rate of primaries as a function of their energy
$R(E_p)=k/E_p^3$
\cite{REF} can be used to calculate the rate $R({\cal E}_{\nu
EW}>{\cal E}_\nu^0)$. The result will certainly depend on the
calibrating factor $s$. Comparing this rate with the experimental
upper bound, $R^{up}({\cal E}_{\nu EW}>{\cal E}_\nu^0)$, an upper
limit will be placed on the calibrating factor.

In the process of evaluating $R({\cal E}_{\nu EW}>{\cal
E}_\nu^0)$ we will  need to integrate the rate of primaries
$R(E_p)$ over solid angle, energy, and detection area. The
question of detection area is of great importance since its lack
of symmetry with respect to the antenna compelled us to make an
important simplifying assumption.

\subsection*{Detection area}

\begin{figure}[t]
\centerline{\includegraphics[width=1\textwidth]{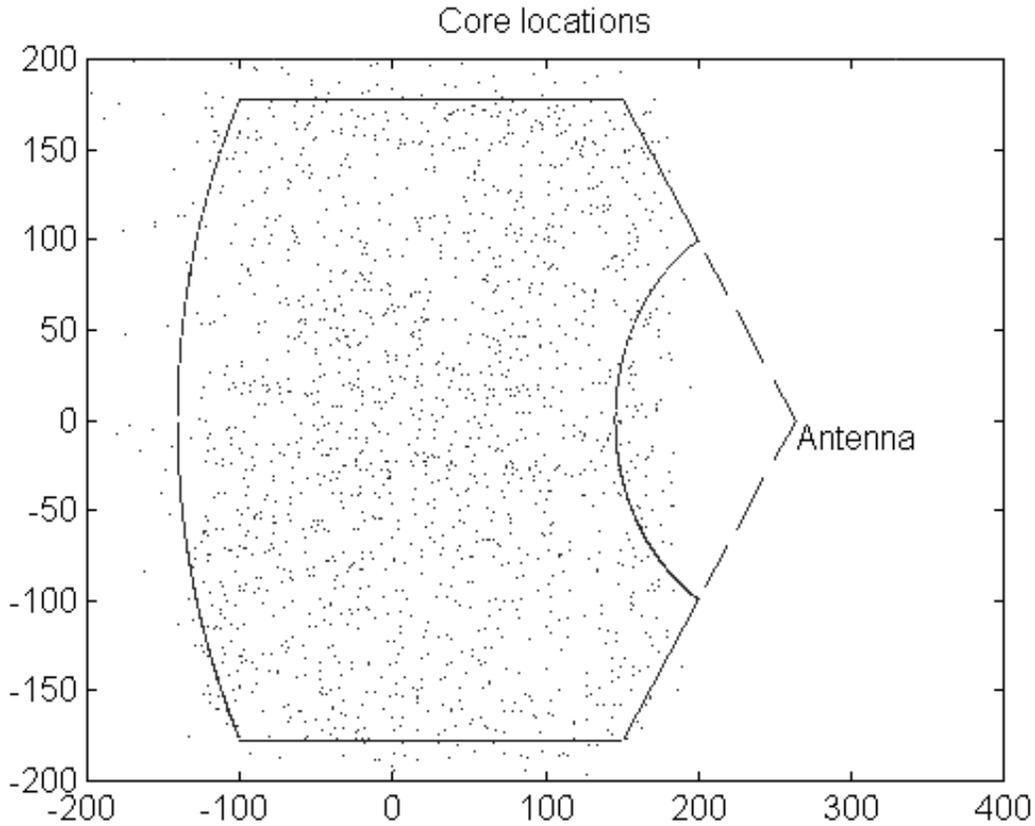}} \caption{Core
locations for 1702 showers giving rise to muon triggering. The
opening angle at the antenna vertex is 2 radians.  The radii of
the inner and outer circular arcs are 100 and 400~m,
respectively.  Axes relative to center of array are $x$
(geographic East) and $y$ (geographic North), in meters.}
\end{figure}

Core locations of showers that fire muon triggering are shown in
Fig.~18.  Their distribution over the outlined area appears to be
approximately uniform. To simplify the integration over solid
angle, area, and energy, let us consider a bigger area: part of
the ring with inner radius of 100 m, outer radius of 400 m,
inside the same angle of 2 radians. The rate of showers passing
through this area is bigger than through the one shown in Fig.~18. 
Integration over the bigger area will lead to a greater value
for $R({\cal E}_{\nu EW}>{\cal E}_\nu^0)$.
However, that value will differ from the one calculated after
integration over the outlined area by less than 5\%, as we show
below in Calculation.

As was mentioned in Section~4.6, systematic studies have been
performed  only for simulated pulses with $\delta=5$~ns. Thus,
$R^{up}({\cal E}_{\nu EW}>{\cal E}_\nu^0)$ [Fig.\ 15(b)] was
obtained under the assumption $R \simeq 200$~m. Though in
reality the detection area is not limited to the area around 
200~m, we are going to accept that result as a reasonable
approximation.

Unfortunately, the area under consideration is not the full ring.
If it were, it could have been proven that the angular distribution
of the electric vector in the horizontal plane is approximately
uniform. Then it would have been easy to derive $R({\cal E}_{\nu
EW}>{\cal E}_\nu^0)$ after calculating $R({\cal E}_\nu>{\cal
E}_\nu^0)$. In that case we wouldn't have to worry about the
polarization of radiation from individual showers.

In our case, however, the area is not symmetric. The general
approach of calculating ${\cal E}_{\nu EW}$ for showers with
different zenith angles, azimuth angles and core locations and
then integrating would be too difficult to implement. Therefore,
it would be reasonable to simplify the problem by making the
assumption that all showers are vertical. This will lead to a
rough estimate of the upper limit that can be placed on the
calibrating factor.

Now we are almost ready to perform integration. Only one thing
is missing: Since $R^{up}({\cal E}_{\nu EW}>{\cal E}_\nu^0)$ was
calculated for the EW projection of the electric field vector, we
will need a relation between ${\cal E}_\nu$ and ${\cal E}_{\nu
EW}$ for showers with different core locations. For this purpose
let us turn to the question of polarization of radiation produced
by a shower.

\subsection*{Polarization}

The conventional approach assumes linear polarization in the
direction perpendicular to both shower axis and magnetic field
vector. The source of such radiation is the transverse current of
shower particles \cite{KL}.
Alternatively, one can consider radio emission
due to particles' {\it acceleration} in the Earth's magnetic field.
For the time being we will pursue the latter approach but at the end
of this Appendix we will give results for both alternatives.

The magnitude and direction of the electric field vector for a
radiating particle are governed by the general formula
$$
{\bf E}({\bf x},t)=e\,\left[\frac{{\bf n}-\boldsymbol{\beta}} {\gamma^2 
(1-\boldsymbol{\beta}\cdot{\bf n})^3\,l^2}\right]_{\rm ret} + 
\frac{e}{c}\,\left[\frac{{\bf n}\times\left[({\bf
n}-\boldsymbol{\beta})\times\dot{\boldsymbol{\beta}}\right]} 
{(1-\boldsymbol{\beta}\cdot{\bf 
n})^3\,l}\right]_{\rm ret} \ \ \ \ \ \ \ (A.1)
$$
where $\boldsymbol{\beta}$ is the velocity vector in the units of $c$,
$\dot{\boldsymbol{\beta}}=d\boldsymbol{\beta}/dt$ is the acceleration vector,
divided by $c$, ${\bf  n}$ is a unit vector from the radiating
particle to the antenna, and $l$ is the distance to the particle~\cite{JDJ}. 
The square brackets with subscript ``ret" mean that
the quantities in the brackets are evaluated at the retarded time. (A.1) does
not consider the influence of the index of refraction of air. Here we are 
primarily interested in the direction of the electric field vector and for 
$R>100$ m the effect of the refraction index on polarization is 
insignificant. Ref.~\cite{SRG} will give a full calculation of shower 
radiation, employing a modified formula (A.1) to take into account the 
refraction index.  

The first term in (A.1) decreases with distance as $1/l^2$ and represents a 
boosted Coulomb field. It does not produce any radiation. 
The magnitudes of two terms in (A.1) are related as 
$1/(\gamma^2 l)$ and $|\dot{\boldsymbol{\beta}}|/c$. 
The characteristic acceleration of a 30~MeV electron ($\gamma\approx60$) of 
an air shower in the Earth's magnetic field 
($B\approx0.5$~Gauss) is $|{\bf a}|=ecB/(\gamma m) \approx 
4.4\cdot10^{13}$~m/s$^2$.
Even when an electron is as close to the antenna as 1000~m,
the first term is three orders of magnitude 
smaller than the second and can be neglected. 
The second term falls as $1/l$ 
and is associated with a radiation field. It describes the electric field of 
a single radiating particle for most geometries relevant to extensive
air showers. 

\begin{figure}[t]
\centerline{\includegraphics[width=.9
\textwidth]{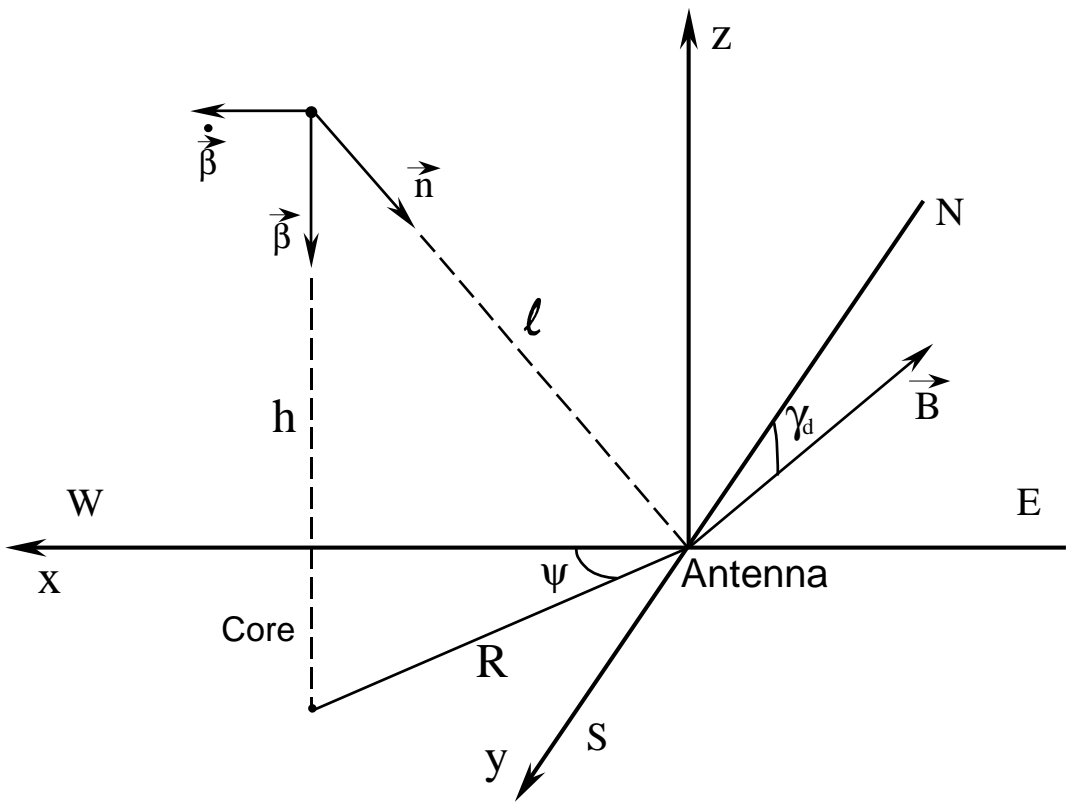}}
\caption{Geometry of a vertical shower. Axes relative to  antenna
are $x$ (magnetic West), $y$ (magnetic South) and $z$ (up). 
Vector {\bf B} lies in the $yOz$ plane.}
\end{figure}

Consider the frame centered at the antenna, with axis $Ox$ going
to  the magnetic West, $Oy$ to the South and $Oz$ directly up. For
vertical showers in this frame $\boldsymbol{\beta}=(0,0,-1)$, while
$\dot{\boldsymbol{\beta}}$ is parallel to $Ox$, or, in other words, to the
$(1,0,0)$ vector (see Fig.~19).\  Let $\psi$ be the angle between
$Ox$ and the direction to the shower core, $R$ the distance to the
core, and $h$ the altitude of the radiating
particle. Then ${\bf n}=\left(-\frac{R\cos\psi }{\sqrt{h^2+R^2}},
-\frac{R\sin\psi}{\sqrt{h^2+R^2}},-\frac{h}{\sqrt{h^2+R^2}}\right)$.
Typical values for $R$ are a few hundred meters, and for $h$
several kilometers, so $R/h$ can be considered small. 
The denominator of the second term of Eq.~(A.1) is independent 
of $\psi$. The numerator determines that, to leading
(second) order in $R/h$, the
electric field vector lies in the horizontal plane and is
parallel to $(\cos2\psi,\sin2\psi,0)$. The magnitude of the numerator is 
independent of the angle $\psi$ up to terms of the order $R^4/h^4$.

This result shows that although particles are accelerated by the
Earth's magnetic field in the EW direction, the polarization of
the resulting radiation does not show preference for a particular
direction. In other words, assuming uniform distribution of
shower cores around the antenna (uniform distribution of $\psi$),
one obtains a uniform angular distribution of the electric field vector.

The most important result of this section for the following
calculation is the relation between ${\cal E}_{\nu EW}$ and
${\cal E}_\nu$:
$$
{\cal E}_{\nu EW}={\cal E}_\nu|\cos2\psi|
$$

\subsection*{Calculation}

Let us now calculate the rate $R({\cal E}_{\nu EW}>{\cal
E}_\nu^0)$ as a function of $s$. For vertical showers Eq.~(1) 
becomes
$$
{\cal E}_\nu=s\,\frac{E_p}{10^{17}~\mbox{eV}}\,
\cos\gamma_d\,\exp\left(-\frac{R}{R_0(\nu,0)}\right)\ \ \ \
\mu \mbox{V m}^{-1} \mbox{MHz}^{-1} \,
$$
where $s$ is the calibrating factor, and $\gamma_d$ is a dip angle (the angle
between the magnetic field vector and the direction to magnetic
North). $R_0(\nu,\theta)$ only depends slightly on $\theta$ when
$\theta<35^{\circ}$. The dependence on $\nu$ is not very
significant, either. Ref.\ [2] quotes two numbers: At
$\nu=55$~MHz $R_0$ equals 110 m, while at $\nu=32$~MHz $R_0$
equals 140 m. The form of the dependence is not well known,
although it is clear that $R_0$ becomes larger as $\nu$
decreases. Assuming linear dependence for simplicity, one can
evaluate $R_0=130$~m at $\nu=39$~MHz, the median frequency of the
main region of investigation [Section~5.3.1 (b)]. This value of
$R_0$ will be used in this Appendix.

Consider showers with impact parameter $R$, direction to the
core  given by angle $\psi$, and primary energy
$$
E_p>E'_p({\cal E}_\nu^0, R, \psi)={\cal
E}_\nu^0\,\frac{10^{17}}{s\cos\gamma_d}\,\frac{\exp(R/R_0)}{|\cos2\psi|}\
\ \ \mbox{eV}
$$
where ${\cal E}_\nu^0$ is expressed in $\mu \mbox{V m}^{-1} \mbox{MHz}^{-1}$.
Only these showers induce ${\cal E}_{\nu
EW}={\cal E}_\nu\,|\cos2\psi|>{\cal E}_\nu^0$. So, the rate of
events where the East-West projection of the pulse is greater
than some value, $R({\cal E}_{\nu EW}>{\cal E}_\nu^0)$, equals
the rate $Z$ of such showers passing through the detection area.

Now let us determine the limits of integration. We assume that
although we consider vertical showers only, the solid angle of
observation is limited by a zenith angle $\theta_m=50^{\circ}$.

We also take into account that magnetic North at Dugway is located
$14^{\circ}$ east of true North. Therefore, the detection area is
bound by lines $\psi=57^{\circ}+14^{\circ}=71^{\circ}$ and
$\psi=-57^{\circ}+14^{\circ}=-43^{\circ}$.

We know that the rate of primaries (in
$\mbox{eV}^{-1}\mbox{km}^{-2}\mbox{sr}^{-1}\mbox{h}^{-1}$)  is
given by $R(E_p)=k/E_p^3$. Hence, the integration over solid
angle, area and energy gives the desired rate $Z$ (in
$\mbox{h}^{-1}$)
$$
\begin{array}{l} Z=\int d\Omega\int dA\int\limits_{E'_p({\cal E}_\nu^0,
R,\psi)}^{+\infty}  R(E_p)\,dE_p=\\ \quad
2\pi\int\limits_0^{\theta_m}\sin\theta\,d\theta
\int\limits_{-43^{\circ}}^{71^{\circ}}d\psi
\int\limits_{100~\mbox{m}}^{400~\mbox{m}}R\,dR\int\limits_{E'_p({\cal
E}_\nu^0, R, \psi)}^{+\infty}\frac{k}{E_p^3}\,dE_p=\\ \quad
2\pi(1-\cos\theta_m)\int\limits_{-43^{\circ}}^{71^{\circ}}d\psi
\int\limits_{100~\mbox{m}}^
{400~\mbox{m}}R\,dR\,\,\frac{k}{2E_p^{\prime2}}=\\\quad\ \pi
k(1-\cos\theta_m)\left(\frac{s\cos\gamma_d}{10^{17}{\cal E}_\nu^0}\right)^2
\int\limits_{-43^{\circ}}^{71^{\circ}}\cos^22\psi\,d\psi
\int\limits_{100~\mbox{m}}^{400~\mbox{m}}Re^{-2R/R_0}\,dR\
\ \ \ \ \ \ (A.2)
\end{array}
$$
Both integrals in Eq.~(A.2) can be readily calculated to
obtain the following expression:
$$
R({\cal E}_{\nu EW}>{\cal E}_\nu^0)=Z\approx\pi
k(1-\cos\theta_m)\left(\frac{s\cos\gamma_d}{10^{17}{\cal
E}_\nu^0}\right)^2\cdot2.0\cdot10^3 \ \ \ \ \
\mbox{h}^{-1}
$$
where $k$ is in $\mbox{eV}^2\mbox{m}^{-2}\mbox{sr}^{-1}\mbox{h}^{-1}$ and
${\cal E}_\nu^0$ is in $\mu\mbox{V/m/MHz}$.
Let us note here that integration over the area outlined in Fig.~18
(instead of the part of the ring) has also been calculated numerically with 
the help
of {\it Mathematica}, with the result
$$ R({\cal E}_{\nu EW}>{\cal
E}_\nu^0)\approx\pi k(1-\cos\theta_m)\left(\frac{s\cos\gamma_d}{10^{17}{\cal
E}_\nu^0}\right)^2\cdot1.9\cdot10^3 \ \ \ \ \
\mbox{h}^{-1}
$$
i.e., only a 5\% difference. Thus, the rapidly decreasing exponential
suppresses the contribution at large distances from the antenna where
the part of the ring does not overlap the outlined area.

Now we can place an upper limit on $s$ using this formula for the
rate $R({\cal E}_{\nu EW}>{\cal E}_\nu^0)$ and the experimental
upper limit on this rate, $R^{up}({\cal E}_{\nu EW}>{\cal
E}_\nu^0)$ from Section~5.3.4:
$$
R({\cal E}_{\nu EW}>{\cal E}_\nu^0) \approx
\pi k(1-\cos\theta_m)\left(\frac{s\cos\gamma_d}{10^{17}{\cal
E}_\nu^0}\right)^2\cdot1.9\cdot10^3<R^{up}({\cal E}_{\nu EW}>{\cal
E}_\nu^0) \ \ \ \ \ (A.3)
$$
$$
s<\frac{10^{17}{\cal
E}_\nu^0}{\cos\gamma_d}\cdot\sqrt{\frac{R^{up}({\cal E}_{\nu
EW}>{\cal E}_\nu^0)}{\pi k(1-\cos\theta_m)\cdot1.9\cdot10^3}} \ \
\ \ \ \ \ \ \ \ \ \ (A.4)
$$
>From Eq.~(\ref{eqn:uplimEW}),
${\cal E}_\nu^0\cdot\sqrt{R^{up}({\cal E}_{\nu EW}>{\cal
E}_\nu^0)}\approx0.943~\mu\mbox{V/m/MHz/h}^{1/2}$. This 
value, substituted into~(A.4) together with $k=3.84\cdot10^{24}~\mbox{eV}^2
\mbox{m}^{-2}\mbox{sr}^{-1}\mbox{s}^{-1} =
1.38\cdot10^{28}~\mbox{eV}^2 \mbox{m}^{-2}\mbox{sr}^{-1}\mbox{h}^{-1} \
$ \cite{REF}, $\theta_m=50^{\circ}$ and $\gamma_d=68^{\circ}$, gives the 
final
result: $s<46$. A similar calculation employing North-South
results [Eq.~(\ref{eqn:uplimNS})] gives $s<34$.

If the main source of radiation is transverse current, not
particles' acceleration, then ${\bf E}$ is parallel to
$\boldsymbol{\beta}\times{\bf B}$~\cite{KL} and for vertical showers ${\cal
E}_{\nu EW}={\cal E}_\nu$.
This changes the integral of $\cos^2 2\psi$ over $\psi$ in formula~(A.2)
to the integral of 1 over $\psi$, leading to $s<31$. This is the
smallest upper limit we can place on $s$.
Unfortunately, this upper limit is not quite small enough to rule
out the initial Haverah Park group claim of $s=20$, let alone the
values of 9.2 or 1.6. To facilitate comparison of results of
different studies, let's rewrite formula~(A.3) under the
assumption of the transverse current mechanism. In simplified form it
is
$$
R({\cal E}_{\nu EW}>{\cal
E}_\nu^0)\approx9.16\cdot10^{-4}\,\left(\frac{s}{{\cal E}_\nu^0}\right)^2
$$

The transverse current mechanism cannot account for the NS component for 
vertical showers so we will employ the model discussed above in Polarization.
According to it, ${\cal E}_{\nu NS}={\cal E}_\nu|\sin2\psi|$. Then, 
formula~(A.3) transforms into 
$$
R({\cal E}_{\nu NS}>{\cal
E}_\nu^0)\approx4.86\cdot10^{-4}\,\left(\frac{s}{{\cal
E}_\nu^0}\right)^2
$$ 
The plots of $R({\cal E}_{\nu EW}>{\cal
E}_\nu^0)$ and $R({\cal E}_{\nu NS}>{\cal
E}_\nu^0)$ for different values of $s$ are shown in Figs.~16 and 17.

The noise level at Dugway is too high and the acquired sample is too
limited in statistics and dynamic range
to allow us to place upper limits strict enough to
check the claims of the two groups. We hope that further
improvement of the data processing technique will reduce the
noise contribution. The dip angle $\gamma_d$ is much smaller at the
Auger site in Argentina ($34^{\circ}$ versus $68^{\circ}$ at
Utah). This leads to bigger electric fields for vertical showers,
facilitating the detection of shower radiation. We also expect
that the Argentina site will be quieter than Dugway and some
clarity as regards the calibrating factor will be established.

\section*{Appendix B: Relation between simulated pulse strength
and peak detected voltage}

Within the bandwidth $\delta\nu$, the antenna feels the electric field
${\cal E}_\nu\cdot\delta\nu$. Then, the instantaneous power flow is 
given by $({\cal E}_\nu\cdot\delta\nu)^2/(120\pi)$, where 
$120\pi~\Omega$ is the impedance of free space. The antenna effective
aperture is equal
to $G_{ant}\,(\lambda^2/4\pi)$, $G_{ant}\simeq2.5$ being the
antenna gain. 
Hence, it supplies a power of
$$
W=\frac{({\cal E}_\nu\cdot\delta\nu)^2}{(120\pi)} \cdot
G_{ant}\,\frac{\lambda^2}{4\pi}
$$
to the input of the feedline cable. 
Subsequently, this power is transmitted by 60 feet of RG-58U cable 
with average attenuation at 39 MHz estimated to be 1.44~dB or a 
factor of $L\simeq1.4$ in power. So, the power at the input of the 
preamplifier is $W/L$. 
The impedances in the antenna system were matched, in particular, the 
resistance of the preamplifier was equal to the characteristic impedance of 
the cable, $R=50~\Omega$. The power at the input of the 
preamplifier is transformed into a peak voltage $V_{pk}$ across the 
preamplifier resistance $R$, i.e.\ $W/L=V_{pk}^2/R$.
Thus,
$$
{\cal E}_\nu=\frac{1}{\delta\nu}\sqrt{W\,\frac{480\pi^2}{G_{ant}\lambda^2}}
= \frac{\nu}{\delta\nu}\,\frac{2V_{pk}}{c}\,\sqrt\frac{120\pi^2L}{G_{ant}R}
= \frac{\nu}{\delta\nu}\,\frac{2V_{pk}}{c}\,\sqrt\frac{120\pi^2}{GR}
$$
where $G=G_{ant}/L\simeq1.8$ is the antenna-cable system gain.
For specific values used in data analysis (Sec.~5.3.1.)  $\delta\nu=54-24=30$~MHz, $\nu=39$~MHz,
$$
{\cal E}_\nu=0.042\,\frac{V_{pk}}{\sqrt{G}}
=0.031\,\frac{V_{pk}}{\sqrt{G/1.8}} \ \ \ \mu\mbox{V/m/MHz}
$$
where $V_{pk}$ is in $\mu$V.

The above derivation implies that the electric field perceived by the
antenna is due to frequencies in the $\delta\nu$ range only.
However, the simulated pulse voltage with $V_{pk}=1.3$~mV has components 
in the whole frequency range (see Fig.~4, bottom), i.e., it is the result 
of an imaginary ``electric field" with very diverse frequencies. To
find what $V_{pk}$ would be if our antenna were sensitive to
24-54~MHz only, the forward and then inverse (for 24-54~MHz
region) Fourier transforms of the simulated pulse voltage have been
performed. The contribution of 24-54~MHz region into the pulse
peak voltage of 1.3~mV appeared to be equal to 442~$\mu$V.
This leads to a corrected formula:
$$
{\cal E}_\nu=0.031\,\frac{442}{1300}\,\frac{V_{pk}}{\sqrt{G/1.8}}=0.011\,
\frac{V_{pk}}{\sqrt{G/1.8}}
\ \ \ \mu\mbox{V/m/MHz}
$$

This formula establishes the relation between the simulated pulse 
strength ${\cal E}_\nu$ and its peak voltage $V_{pk}$ at 
the filter-preamplifier configuration. 
$G=1.8$ was assumed throughout the paper.

\section*{Appendix C: Cost considerations for Auger project}

At present we can only present a rough sketch of criteria for
detection in the 30--100~MHz range. Data would be digitized at a
500~MHz rate at each station and stored in a rolling manner, with
at least 20 microseconds of data in the pipeline at any moment.
Upon receipt of a trigger signaling the presence of a ``large''
shower ($> 10^{18}$~eV), these data would be merged into the rest
of the data stream at each station.

Per station, we estimate the following additional costs, in US
dollars, for RF pulse detection:

\begin{center}
\begin{tabular}{l l c} \hline
Two antennas and protection circuitry:   & 200 & (a) \\
Mounting hardware:                       & 100 & (b) \\
Cables and connectors:                   & 200 & (c) \\
Preamps:                                 & 500 & (d) \\
Digitization and memory electronics:     &2000 & (e) \\ \hline
Total per station:                       &3000 & (f) \\ \hline
\end{tabular}
\end{center}

\noindent

\noindent (a) Two military-surplus log-periodic antennas; crossed
polarizations.

\noindent (b) Highly dependent on other installations at site.
Antennas are to be pointed vertically but optimum elevation not
yet determined.

\noindent (c) Antennas are mounted near central data acquisition
site of each station, but sufficiently far from any sources of RF
interference such as switching power supplies. Alternative
location of receiving stations at interstitial positions in the
array would require microwave communication links and auxiliary
power supplies and would add to cost.

\noindent (d) Commercial GaAsFET preamps and gas discharge tubes.

\noindent (e) Subject to prototype development experience.  Power
requirements not yet known.

\noindent (f) The number of stations to be equipped with RF
detection will depend on further prototype experience.
\bigskip

The above estimate assumes that one can power the preamps and DAQ
electronics from the supply at each station without substantial
added cost.  It also assumes that a ``large-event trigger'' will
be available at each station.
One consideration may be the acquisition of antennas
robust enough to withstand extreme weather (particularly wind)
conditions.

For detection at frequencies above or below 30--100~MHz, the
criteria are not yet well enough developed to permit any cost
estimate.

\end{document}